\documentclass[final,3p,times,twocolumn]{elsarticle}

\usepackage{amssymb}
\usepackage{amsbsy}
\usepackage{verbatim}

\journal{Chemical Physics}

\begin{document}

\begin{frontmatter}



\title{Effect of a high-frequency magnetic field on the resonant behavior displayed by a spin-$1/2$ particle under the influence of a rotating magnetic field}


\author{Jes\'us Casado-Pascual}

\address{F\'{\i}sica Te\'orica, Universidad de Sevilla,\\ 
        Apartado de Correos 1065, Sevilla 41080, Spain}

\begin{abstract}
In this paper, we investigate the role of a high-frequency magnetic field in the resonant behavior displayed by a spin-$1/2$ particle under the influence of a rotating magnetic field. We propose two alternative methods for analyzing the system dynamics, namely, the averaging method and the multiple scale method. The analytical results achieved by applying these two methods are compared with those obtained from the numerical solution of the Schr\"odinger equation. This comparison leads to the conclusion that the multiple scale method provides a better understanding of the system dynamics than the averaging method. In particular, the averaging method predicts the complete destruction of the resonant behavior by an appropriate choice of the parameter values of the high-frequency magnetic field. This conclusion is disproved both by the numerical results, and also by the results obtained from the multiple scale method. 
\end{abstract}

\begin{keyword}
Coherent destruction of tunneling \sep resonant behavior \sep spin-$1/2$ particle \sep rotating magnetic field \sep high-frequency fields \sep two-state systems


\end{keyword}

\end{frontmatter}


\section{Introduction}
\label{Introduction}
The dynamics of quantum systems under the influence of time-periodic driving fields
is a long-standing problem in Quantum Mechanics \cite{Schwinger,Rabi,Bloch,Torrey,Autler}. An extensive review on this topic, covering both non-dissipative and dissipative systems, can be found in Ref.~\cite{Grifoni&Hanggi}. For a more concise review with an accurate description of the Floquet-theoretical method, it is recommended to consult chapter 5 in Ref.~\cite{Hanggi1}. 

In the past two decades, the study of driven quantum systems has sparked renewed interest after the discovery of the phenomenon of coherent destruction of tunneling \cite{Grossmann}. In the original work~\cite{Grossmann}, the authors considered the problem of a particle moving in a double-well potential and perturbed by a monochromatic driving field.  The analysis of the Floquet spectrum of the system allowed them to show that, by an appropriate choice of the parameter values of the driving field, the tunneling of the particle through the potential barrier can be brought to a standstill. Among the possible applications of this phenomenon, one can mention the laser-induced trapping of an electron in a
quantum-well structure \cite{Bavli1,Bavli2}, and the control of proton or electron transfer reactions \cite{Morillo,Dakhnovskii,Grifoni&Hanggi2}. Recently, the phenomenon of coherent destruction of tunneling has been experimentally verified in an optical double well system \cite{DellaValle}, in a Bose-Einstein condensate in a strongly driven optical lattice \cite{Lignier} and in  
single-particle tunneling in strongly driven double-well potentials \cite{Kierig}.

In order to clarify the mechanism of coherent destruction of tunneling reported in Ref.~\cite{Grossmann}, it has turned out to be useful to consider a simplified two-state model \cite{Grossmann1,Gomez-LLorente,Wang,Cheng,Kayanuma1,Zhao,Creffield1,Creffield2}. In this model, all the spatial information contained in the Floquet modes is neglected, and only the influence of the lowest quasienergy doublet is taken into account. The study of this approximate two-state model makes it possible to determine simple conditions for the appearance of the phenomenon of coherent destruction of tunneling. For instance, in the high-frequency limit, an averaging procedure shows that this phenomenon occurs when the ratio of the amplitude of the monochromatic driving field to its frequency is a zero of the zeroth-order Bessel function of the first kind \cite{Kayanuma1}. It is worthwhile to point out that the phenomenon of coherent destruction of tunneling in a two-state model resembles that of dynamic localization observed in a driven, infinite tight-binding system \cite{Dunlap}. A discussion of their similarities and differences can be found in Sec.~4 of Ref.~\cite{Grifoni&Hanggi}. Recently, the internal relationship between these two phenomena has been clarified in Ref.~\cite{Kayanuma}. 

In the last decade, several works have been published which describe alternative approaches, beyond the averaging procedure, for the analytical study of a driven two-state quantum system \cite{Barata1,Barata2,Barata3,Frasca1,Frasca2,Xie}. In particular, in Refs.~\cite{Barata1,Barata2,Barata3} the authors propose a convergent strong-coupling expansion free of secular terms. In Refs.~\cite{Frasca1,Frasca2} the method is based on dual Dyson series and renormalization techniques. Finally, in Ref.~\cite{Xie} the author applies a multiple scale expansion in the inverse of the driving frequency for the calculation of the Floquet states and the quasienergies. One of the main conclusions of these works is that, for high but finite values of the driving frequency, the coherent destruction of tunneling in a two-state quantum system is only temporary. More precisely, in the high-frequency limit, coherent destruction of tunneling occurs only for time values much smaller than a characteristic time scale proportional to the driving frequency squared. This is a consequence of the fact that the quasienergy degeneracy predicted by the averaging procedure when the ratio of the amplitude of the monochromatic driving field to its frequency is a zero of the zeroth-order Bessel function of the first kind is exact only in the mathematical limit of infinite driving frequency. For high but finite values of the driving frequency, there is a quasienergy splitting of the order of the inverse of the driving frequency squared (see, for instance, Ref.~\cite{Xie}).

In the present paper, we take up again the problem of a spin-$1/2$ particle in the presence of a rotating magnetic field, originally studied by Rabi~\cite{Rabi}. This same model has been considered in Ref.~\cite{Shao¬Hanggi} to study the suppression of quantum coherence induced by circularly polarized driving fields. As is well known, under the influence of a rotating magnetic field, the dynamics of a spin-$1/2$ particle displays a resonant behavior (see, for instance, Sec.~11.2 of Ref.~\cite{Galindo} or the beginning of Sec.~\ref{Results} of the present paper). In this work, we are interested in analyzing how this resonant behavior is affected by the application of an additional high-frequency magnetic field along the rotation axis. As will be shown in this paper,  
there is a remarkable similarity between the problem considered here and the problem of coherent destruction of tunneling in a two-state quantum system driven by a monochromatic field.

The paper is organized as follows. In the following section, we introduce the model and apply an averaging method to work out an analytical expression for the time evolution operator. In Sec.~\ref{MSsection}, we use multiple scale techniques \cite{Bender&Orszag,Janowicz} to obtain an improved expression for the time evolution operator which is valid in a wider range of time scales than the one obtained by the previous averaging procedure. In Sec.~\ref{Results}, we apply the analytical results in the previous sections to the study of the effects of the high-frequency magnetic field on the dynamics of the system. In particular, we examine in detail its effects on the above mentioned resonant behavior. In that section, we also compare our analytical results with those obtained from the numerical solution of the Schr\"odinger equation.  Finally, in Sec.~\ref{Conclusions}, we present conclusions for the main findings of our work.

\section{Description of the model. The averaging method}
\label{description}

\begin{figure}[htb]
\includegraphics[scale=0.6]{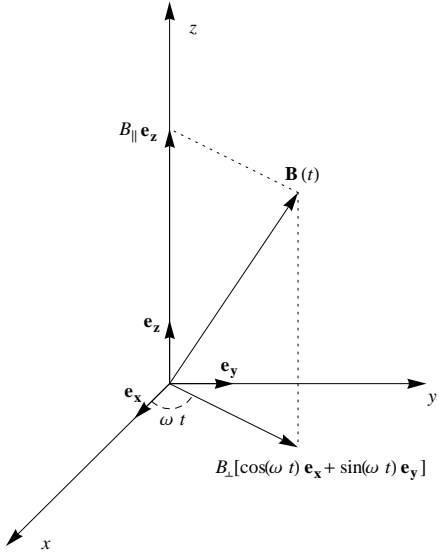}
\caption{Sketch of the gyrating magnetic field in Eq.~(\ref{MF}).}
\label{field}
\end{figure}

The effect of a uniform, gyrating magnetic field of the form 
\begin{equation}
\label{MF}
\mathbf{B}(t)=B _{\perp}\left[\cos(\omega t) \mathbf{e _ x}+\sin(\omega t) \mathbf{e _ y}\right]+B _{\parallel}\mathbf{e _ z}
\end{equation}
on the dynamics of a spin-$1/2$ particle has been well-known since the classic paper by Rabi \cite{Rabi}. In the above expression $\mathbf{e _ x}$, $\mathbf{e _ y}$ and $\mathbf{e _ z}$ are the 3-D unit vectors along the $x$, $y$ and $z$ axes, respectively, the constants $B _{\parallel}$ and $B _{\perp}$ are the components of the magnetic field parallel and perpendicular to the $z$ axis, respectively, and $\omega$ is the angular frequency of the magnetic field's rotation [see Fig.~(\ref{field})]. 

The time evolution of a state vector $|\Psi,t\rangle$ is governed by the Schr\"odinger equation
\begin{equation}
\label{Schrodingerequation}
i  \hbar \frac{\partial|\Psi,t\rangle}{\partial t}=-\gamma \mathbf{B}(t)\cdot \hat{\mathbf{S}} |\Psi,t\rangle, 
\end{equation}  
where $\gamma$ is the gyromagnetic ratio, $\hat{\mathbf{S}}=\hbar(\hat{\sigma}_x \mathbf{e _ x}+\hat{\sigma}_y \mathbf{e _ y}+\hat{\sigma}_z \mathbf{e _ z})/2$ is the 3-D spin operator (with $\hat{\sigma}_x$, $\hat{\sigma}_y$ and $\hat{\sigma}_z$ being the standard Pauli matrices), and the centered dot denotes the usual scalar product of two 3-D vectors. 

One of the main features of this system is the fact that, under the influence of the rotating field in Eq.~(\ref{MF}), the spin dynamics displays a resonant behavior.  This resonance behavior manifests itself in several ways. For instance, as discussed in Sec.~\ref{Results}, if one takes as initial condition $|\Psi,0\rangle$ an eigenstate of $\hat{\sigma}_z$ and assumes that $B_{\perp}\neq 0$, the expected value $\langle \hat{\sigma}_z,t\rangle=\langle \Psi,t|\hat{\sigma}_z|\Psi,t\rangle$ is an oscillating function of time with maximum amplitude at an angular frequency $\omega=-\gamma B_{\parallel}$. 

The aim of this paper is to study how this resonant behavior is affected by the presence of an additional high-frequency magnetic field of the form
\begin{equation}
\label{HFF}
\mathbf{B}_{\mathrm{HF}}(t)=B _{\mathrm{HF}} \cos \left( \Omega _{\mathrm{HF}} t+\varphi\right)\mathbf{e _ z},
\end{equation}
where $B _{\mathrm{HF}}$ is the amplitude of the high-frequency field, $\Omega _{\mathrm{HF}}$ is its frequency, and $\varphi\in[0,2 \pi)$ is an arbitrary phase shift between  $\mathbf{B}_{\mathrm{HF}}(t)$ and $\mathbf{B}(t)$. In this case, the equation of motion of the system is given by Eq.~(\ref{Schrodingerequation}) with $\mathbf{B}(t)$ replaced by $\mathbf{B}(t)+\mathbf{B}_{\mathrm{HF}}(t)$. By high-frequency we mean that $\Omega _{\mathrm{HF}}$ is much higher than the rest of the characteristic frequencies of the problem in the absence of $\mathbf{B}_{\mathrm{HF}}(t)$. Thus, if we introduce the characteristic frequencies $\omega_{\parallel}=\gamma B_{\parallel}$ and $\omega_{\perp}=\gamma B_{\perp}$, it is assumed that $\omega_{\perp},\,\omega_{\parallel},\,\omega \ll \Omega _{\mathrm{HF}}$. In principle, the frequency $\omega_{\mathrm{HF}}=\gamma B _{\mathrm{HF}}$ may be of the same order of magnitude as $\Omega _{\mathrm{HF}}$. In what follows, $\omega$ will be taken as the unit of frequency (i.e., $\omega=1$), so that  $\omega t = t$, $\omega_{\perp}/\omega= \omega_{\perp}$, $\omega_{\parallel}/\omega= \omega_{\parallel}$, $\omega_{\mathrm{HF}}/\omega=\omega_{\mathrm{HF}}$, and $\Omega_{\mathrm{HF}}/\omega=\Omega_{\mathrm{HF}}$ are all dimensionless quantities. 

In this section and the following one, we are interested in obtaining analytical expressions for the time evolution operator from time $t=0$ to time $t$. This operator, which will be denoted by $\hat{U}(t)$, fulfills the differential equation
\begin{equation}
\label{evolutionoperator1}
i \frac{\partial \hat{U}(t)}{\partial t} = \hat{H}(t)\hat{U}(t)
\end{equation}
with the initial condition
\begin{equation}
\label{IC}
\hat{U}(0)=\hat{I},
\end{equation}
where 
\begin{eqnarray}
\label{Hamiltonian}
\hat{H}(t)&=&-\frac{\omega_{\perp}}{2}\left[\cos (t) \hat{\sigma}_ x+\sin (t) \hat{\sigma}_ y\right]\nonumber\\
&&-\frac{1}{2}\left[\omega _{\parallel}+\omega_{\mathrm{HF}}\cos \left( \Omega _{\mathrm{HF}} t+\varphi\right)\right]\hat{\sigma}_ z
\end{eqnarray}
is the Hamiltonian operator in dimensionless units, and $\hat{I}$ the identity operator. 

 The differential equation (\ref{evolutionoperator1}) can be transformed to a more convenient form for the calculations by performing the following transformation
\begin{equation}
\label{UTransformation}
\hat{G}(t)=e^{i \left[t -\theta(t)\right]\hat{\sigma}_ z/2}\hat{U}(t),
\end{equation}
with 
\begin{equation}
\label{defthetadet}
\theta(t)=r\sin(\Omega_{\mathrm{HF}} t+\varphi),
\end{equation}
and $r=\omega_{\mathrm{HF}}/\Omega_{\mathrm{HF}}$. Then, it is straightforward to see that the transformed time evolution operator $\hat{G}(t)$ can be obtained by solving the differential equation
\begin{equation}
\label{evolutionoperator2}
i \frac{\partial \hat{G}(t)}{\partial t}=\hat{h}(\Omega_{\mathrm{HF}} t) \hat{G}(t),
\end{equation}
with the initial condition
\begin{equation}
\label{IC2}
\hat{G}(0)=e^{-i \theta(0)\hat{\sigma}_ z/2},
\end{equation}
and the transformed Hamiltonian
\begin{eqnarray}
\label{THamiltonian}
\hat{h}(\Omega_{\mathrm{HF}} t)&=&-\frac{\omega_{\perp}}{2} \{\cos\left[\theta(t)\right]\hat{\sigma}_x+\sin\left[\theta(t)\right] \hat{\sigma}_y\} \nonumber \\ &&-\frac{1+\omega_{\parallel}}{2}\hat{\sigma}_z.
\end{eqnarray}

For the multiple scale method developed in the next section Eq.~(\ref{evolutionoperator2}) offers two advantages over Eq.~(\ref{evolutionoperator1}). First, in contrast to the original Hamiltonian $\hat{H}(t)$, the transformed Hamiltonian $ \hat{h}(\Omega_{\mathrm{HF}} t)$ does not contain terms proportional to $\omega_{\mathrm{HF}}$ which, as mentioned above, may be of the same order of magnitude as the high-frequency $\Omega _{\mathrm{HF}}\gg 1$. Second, as we have made explicit by using the notation $\hat{h}(\Omega_{\mathrm{HF}} t)$, the transformed Hamiltonian depends on time only through $\Omega_{\mathrm{HF}} t$ which, as we will see in Sec.~\ref{MSsection}, is the rapid time scale $t_0$ in the multiple scale analysis.

In the absence of the high-frequency field (i.e., for $r=0$), Eq.~(\ref{evolutionoperator2}) can be solved exactly since the transformed Hamiltonian in Eq.~(\ref{THamiltonian}) becomes time independent. This is not the case for $r\neq 0$ and, consequently,  one has to resort to approximate techniques in order to obtain an analytical solution of Eq.~(\ref{evolutionoperator2}). In this section, we will apply 
an averaging method to solve Eq.~(\ref{evolutionoperator2}). From Eqs.~(\ref{evolutionoperator2}) and (\ref{THamiltonian}) it follows that the time derivative of $\hat{G}(t)$ is at most of order 1 in the  high-frequency limit $\Omega_{\mathrm{HF}}\gg 1$. Therefore, a large number of oscillations of the function $\theta(t)$ appearing in the definition of $\hat{h}(\Omega_{\mathrm{HF}} t)$ 
takes place before a significant change in $\hat{G}(t)$ occurs. As a consequence, for $\Omega _{\mathrm{HF}}\gg 1$, the time dependent Hamiltonian $\hat{h}(\Omega_{\mathrm{HF}} t)$ appearing in Eq.~(\ref{evolutionoperator2}) can be approximated by its average over one high-frequency period. The result is
\begin{equation}
\label{evolutionoperator2average}
i \frac{\partial \hat{G}(t)}{\partial t}=\hat{h}_{\mathrm{eff}}\, \hat{G}(t),
\end{equation}
where $\hat{h}_{\mathrm{eff}}$ is the effective Hamiltonian 
\begin{eqnarray}
\label{effHamiltonian}
\hat{h}_{\mathrm{eff}}&=&\frac{1}{2 \pi} \int_{0}^{2 \pi} \mathrm{d}t_0\,\hat{h}(t_0)\nonumber\\
&=&-\frac{1}{2} \left[\omega_{\perp} J_0(r)\hat{\sigma}_x+(1+\omega_{\parallel})\hat{\sigma}_z\right],
\end{eqnarray}
with $J_0(r)$ being the zeroth-order Bessel function of the first kind \cite{Abramowitz}. Notice that the above result is identical to that obtained in the absence of the high-frequency magnetic field [see Eq.~(\ref{defh0})], but with the perpendicular frequency $\omega_{\perp}$ renormalized to a value $\omega_{\perp} J_0(r)$. The solution of Eq.~(\ref{evolutionoperator2average}) with the initial condition (\ref{IC2}) is 
\begin{equation}
\label{AMsolution1}
\hat{G}(t)=e^{-i t\hat{h}_{\mathrm{eff}}}e^{-i \theta(0)\hat{\sigma}_ z/2}.
\end{equation}

Finally, according to Eq.~(\ref{UTransformation}), the time evolution operator obtained by the average method is
\begin{equation}
\label{AMsolution2}
\hat{U}_{\mathrm{av}}(t)=e^{-i \left[t -\theta(t)\right]\hat{\sigma}_ z/2}e^{-i t\hat{h}_{\mathrm{eff}}}e^{-i \theta(0)\hat{\sigma}_ z/2}.
\end{equation}
 
\section{Multiple scale method}
\label{MSsection}

 In this section, we will use a multiple scale method to obtain an approximate solution of Eq.~(\ref{evolutionoperator2}) valid in a wider range of time scales than the one obtained by the averaging procedure [see Eq.~(\ref{AMsolution2})]. As will be seen in Sec.~\ref{Results}, this method provides a better understanding of the system dynamics than the averaging method, above all in the long-time limit. Since we are interested in the high-frequency limit $\Omega_{\mathrm{HF}}\gg 1$, we will choose  $\epsilon=\Omega_{\mathrm{HF}}^{-1}\ll 1$ as the small dimensionless parameter in the multiple scale asymptotic expansions. 

In order to apply the multiple scale method, we will introduce the rapid time scale  
\begin{equation}
\label{timescale0}
t_0=\Omega_{\mathrm{HF}} t,
\end{equation}
as well as the  time scales 
\begin{equation}
\label{timescales}
t_k=\epsilon^k t_0=\epsilon^{k-1} t,
\end{equation}
with $k=1,\,2\,,\dots$. Notice that $t_1=t$ is the time scale in which the rotating magnetic field evolves. For ease of notation, a function $f$ of all the time scales $t_k$, with $k\geq n$, will be denoted by $f(\mathcal{T}_n)$, [i.e., $f(\mathcal{T}_n)=f(t_n,t_{n+1},\dots)$].

Even though the exact solution of Eq.~(\ref{evolutionoperator2}), $\hat{G}(t)$, is a function of $t$ alone, we will seek solutions of this equation which are functions of all the time scales in Eqs.~(\ref{timescale0}) and (\ref{timescales}) treated as independent variables. In other words, we will seek solutions of Eq.~(\ref{evolutionoperator2}) of the form $\hat{\mathcal{G}}( \mathcal{T}_0)$. This is just an artifice to remove secular behaviors and, at the end of the calculations, all these time scales will be expressed in terms of $t$ by using Eqs.~(\ref{timescale0}) and (\ref{timescales}). 

Making formal use of the chain rule for partial differentiation in Eq.~(\ref{evolutionoperator2}), it is easy to see that the operator $\hat{\mathcal{G}}( \mathcal{T}_0)$ fulfills the differential equation
\begin{equation}
\label{mseq1}
i \sum_{k=0}^{\infty} \epsilon^k\frac{\partial \hat{\mathcal{G}}( \mathcal{T}_0)}{\partial t_k}=\epsilon\, \hat{h}(t_0) \hat{\mathcal{G}}( \mathcal{T}_0).
\end{equation}

If one assumes a perturbation expansion of the form
\begin{equation}
\label{series}
\hat{\mathcal{G}}(\mathcal{T}_0)=\sum_{n=0}^{\infty} \epsilon^n\hat{\mathcal{G}}^{(n)}(\mathcal{T}_0),
\end{equation}
replaces it in Eq.~(\ref{mseq1}), and equates the terms with the same powers of $\epsilon$, one obtains the hierarchy of equations
\begin{equation}
\label{hierarchy0}
i \frac{\partial \hat{\mathcal{G}}^{(0)}(\mathcal{T}_0)}{\partial t_0}=0,
\end{equation}
for $n=0$, and
\begin{equation}
\label{hierarchyn}
i \sum_{k=0}^{n} \frac{\partial \hat{\mathcal{G}}^{(n-k)}(\mathcal{T}_0)}{\partial t_k}=\hat{h}(t_0)\hat{\mathcal{G}}^{(n-1)}(\mathcal{T}_0),
\end{equation}
for $n\geq 1$. This hierarchy of equations will be the starting point of the rest of our analysis.

\subsection{Time evolution operator for time scales $t \ll \epsilon^{-1}$}
\label{MS1}

According to Eq.~(\ref{hierarchy0}), $\hat{\mathcal{G}}^{(0)}(\mathcal{T}_0)$ does not depend on the rapid time scale $t_0$, i.e.,
\begin{equation}
\label{not0}
\hat{\mathcal{G}}^{(0)}(\mathcal{T}_0)=\hat{\mathcal{G}}_{1}^{(0)}(\mathcal{T}_1),
\end{equation}
where $\hat{\mathcal{G}}_{1}^{(0)}(\mathcal{T}_1)$ is an arbitrary function of all the time scales except $t_0$. 

Thus, setting $n=1$ in Eq.~(\ref{hierarchyn}), one obtains
\begin{equation}
i \frac{\partial \hat{\mathcal{G}}^{(1)}(\mathcal{T}_0)}{\partial t_0}=-i \frac{\partial \hat{\mathcal{G}}_{1}^{(0)}(\mathcal{T}_1)}{\partial t_1}+\hat{h}(t_0)\hat{\mathcal{G}}_{1}^{(0)}(\mathcal{T}_1),
\end{equation}
whose solution is
\begin{eqnarray}
\label{res1}
\hat{\mathcal{G}}^{(1)}(\mathcal{T}_0)&=&\hat{\mathcal{G}}_{1}^{(1)}(\mathcal{T}_1)-t_0 \frac{\partial \hat{\mathcal{G}}_{1}^{(0)}(\mathcal{T}_1)}{\partial t_1}\nonumber \\
&&-i \int_{0}^{t_0} \mathrm{d}t_0^{\prime}\,\hat{h}(t_0^{\prime})\,\hat{\mathcal{G}}_{1}^{(0)}(\mathcal{T}_1),
\end{eqnarray}
with $\hat{\mathcal{G}}_{1}^{(1)}(\mathcal{T}_1)$ being another arbitrary function of all the time scales except $t_0$.

In order to highlight the secular terms appearing in the above expression (i.e., the terms which diverge as $t_0\rightarrow +\infty$), it is convenient to introduce the operator 
\begin{equation}
\label{sigmadef}
\hat{\Sigma}(t_0)=\int_{0}^{t_0} \mathrm{d}t_0^{\prime}\left[\hat{h}(t_0^{\prime})-\hat{h}_{\mathrm{eff}}\right],
\end{equation}
where $\hat{h}_{\mathrm{eff}}$ is the effective Hamiltonian defined in Eq.~(\ref{effHamiltonian}). Then, Eq.~(\ref{res1}) can be rewritten in the form
\begin{eqnarray}
\label{res2}
\hat{\mathcal{G}}^{(1)}(\mathcal{T}_0)\!\!\!\!&=&\!\!\!\!\hat{\mathcal{G}}_{1}^{(1)}(\mathcal{T}_1)-i\, \hat{\Sigma}(t_0)\,\hat{\mathcal{G}}_{1}^{(0)}(\mathcal{T}_1) \nonumber \\
\!\!\!\!&&\!\!\!\!-t_0 \left[\frac{\partial \hat{\mathcal{G}}_{1}^{(0)}(\mathcal{T}_1)}{\partial t_1}+i\,\hat{h}_{\mathrm{eff}}\,\hat{\mathcal{G}}_{1}^{(0)}(\mathcal{T}_1)\right]. \nonumber\\
\!\!\!\!&&\!\!\!\!
\end{eqnarray}

From its definition, it is easy to see that $\hat{\Sigma}(t_0)$ is a periodic function of $t_0$ of period $2 \pi$ and, consequently, remains bounded in the limit $t_0\rightarrow +\infty$. Thus, in order to avoid the secular term appearing in Eq.~(\ref{res2}), it is required that
\begin{equation}
\label{res3}
\frac{\partial \hat{\mathcal{G}}_{1}^{(0)}(\mathcal{T}_1)}{\partial t_1}=-i\,\hat{h}_{\mathrm{eff}}\,\hat{\mathcal{G}}_{1}^{(0)}(\mathcal{T}_1).
\end{equation}
The solution of the above equation is
\begin{equation}
\label{resultado1}
\hat{\mathcal{G}}_{1}^{(0)}(\mathcal{T}_1)=e^{-i t_1 \hat{h}_{\mathrm{eff}}}\hat{\mathcal{G}}_{2}^{(0)}(\mathcal{T}_2),
\end{equation}
where $\hat{\mathcal{G}}_{2}^{(0)}(\mathcal{T}_2)$ is an arbitrary function of all the time scales $t_k$ with $k\geq2$. With this choice of $\hat{\mathcal{G}}_{1}^{(0)}(\mathcal{T}_1)$, the secular term appearing in Eq.~(\ref{res2}) vanishes, so that $\hat{\mathcal{G}}^{(1)}(\mathcal{T}_0)$ is given by
\begin{equation}
\label{resultado2}
\hat{\mathcal{G}}^{(1)}(\mathcal{T}_0)=\hat{\mathcal{G}}_{1}^{(1)}(\mathcal{T}_1)-i\, \hat{\Sigma}(t_0)\,\hat{\mathcal{G}}_{1}^{(0)}(\mathcal{T}_1).
\end{equation}

Now we apply the result given in Eq.~(\ref{resultado1}) to time values such that $t\ll \epsilon^{-1}$. In this case, since $t_k\ll 1$ for $k\geq 2$, the function $\hat{\mathcal{G}}_{2}^{(0)}(\mathcal{T}_2)$ appearing in Eq.~(\ref{resultado1}) can be replaced by a constant to be determined from the initial condition in Eq.~(\ref{IC2}). Then, making $t_1=t$ in Eq.~(\ref{resultado1}) and taking into account Eqs.~(\ref{series}) and (\ref{not0}), it results that
\begin{equation}
\label{solution1}
\hat{G}(t)= e^{-i t \hat{h}_{\mathrm{eff}}}e^{-i \theta(0)\hat{\sigma}_ z/2}+\mathcal{O}(\epsilon).
\end{equation}

This last approximate expression coincides with that obtained in Sec.~\ref{description} by using the averaging method. As we will see in the following subsections, the multiple scale method allows us to go beyond this result and obtain approximate expressions of the evolution operator valid in a wider range of time scales.

\subsection{Time evolution operator for time scales $t \ll \epsilon^{-2}$}

Setting $n=2$ in Eq.~(\ref{hierarchyn}) and using Eqs.~(\ref{not0}), (\ref{resultado1}) and (\ref{resultado2}), one obtains
\begin{eqnarray}
\label{order2}
i \,\frac{\partial \hat{\mathcal{G}}^{(2)}(\mathcal{T}_0)}{\partial t_0}&=&-\,i\,\hat{\mathcal{F}}(\mathcal{T}_1)+\,i\,\hat{\Pi}(t_0)\,\hat{\mathcal{G}}_{1}^{(0)}(\mathcal{T}_1)\nonumber\\
&& +\hat{h}(t_0)\hat{\mathcal{G}}_{1}^{(1)}(\mathcal{T}_1),
\end{eqnarray}
where
\begin{equation}
\label{defF}
\hat{\mathcal{F}}(\mathcal{T}_1)=\frac{\partial\hat{\mathcal{G}}_{1}^{(1)}(\mathcal{T}_1)}{\partial t_1}+\frac{\partial \hat{\mathcal{G}}_{1}^{(0)}(\mathcal{T}_1)}{\partial t_2 }
\end{equation}
and
\begin{equation}
\label{Pidef}
\hat{\Pi}(t_0)=\hat{\Sigma}(t_0)\hat{h}_{\mathrm{eff}}-\hat{h}(t_0)\hat{\Sigma}(t_0).
\end{equation}
The integration of Eq.~(\ref{order2}) results in
\begin{eqnarray}
\label{integrated1}
\hat{\mathcal{G}}^{(2)}(\mathcal{T}_0)\!\!\!&=&\!\!\!\hat{\mathcal{G}}^{(2)}_{1}(\mathcal{T}_1) +\,\int_{0}^{t_0}\mathrm{d}t_0^{\prime}\,\left[\hat{\Pi}(t_0^{\prime})\,\hat{\mathcal{G}}_{1}^{(0)}(\mathcal{T}_1)\right.\nonumber\\
\!\!\!&&\!\!\!\left.-\,i\,\hat{h}(t_0^{\prime})\,\hat{\mathcal{G}}_{1}^{(1)}(\mathcal{T}_1)\right]-\,t_0 \hat{\mathcal{F}}(\mathcal{T}_1),
\end{eqnarray}
with $\hat{\mathcal{G}}_{1}^{(2)}(\mathcal{T}_1)$ being an arbitrary function of all the time scales except $t_0$. Introducing the operators
\begin{equation}
\label{Pieffdef}
\hat{\Pi}_{\mathrm{eff}}=\frac{1}{2 \pi} \int_{0}^{2 \pi} \mathrm{d}t_0\,\hat{\Pi}(t_0)
\end{equation} 
and
\begin{equation}
\label{Deltadef}
\hat{\Delta}(t_0)=\int_{0}^{t_0} \mathrm{d}t_0^{\prime}\,\left[\hat{\Pi}(t_0^{\prime})-\hat{\Pi}_{\mathrm{eff}}\right],
\end{equation}
and using Eq.~(\ref{sigmadef}), Eq.~(\ref{integrated1}) can be rewritten in the form
\begin{eqnarray}
\label{resultado3}
\hat{\mathcal{G}}^{(2)}(\mathcal{T}_0)\!\!\!\!&=&\!\!\!\!\hat{\mathcal{G}}^{(2)}_{1}(\mathcal{T}_1)-\,t_0 \hat{\mathcal{F}}^{\prime}(\mathcal{T}_1)+\,\hat{\Delta}(t_0)\,\hat{\mathcal{G}}_{1}^{(0)}(\mathcal{T}_1)\nonumber\\
\!\!\!\!&& \!\!\!\!-i\,\hat{\Sigma}(t_0)\,\hat{\mathcal{G}}_{1}^{(1)}(\mathcal{T}_1),
\end{eqnarray}
with
\begin{eqnarray}
\hat{\mathcal{F}}^{\prime}(\mathcal{T}_1)&=&\hat{\mathcal{F}}(\mathcal{T}_1)-\,\hat{\Pi}_{\mathrm{eff}}\,\hat{\mathcal{G}}_{1}^{(0)}(\mathcal{T}_1)\nonumber\\
&&+\,i\,\hat{h}_{\mathrm{eff}}\,\hat{\mathcal{G}}_{1}^{(1)}(\mathcal{T}_1).
\end{eqnarray}

As in the case of the operator $\hat{\Sigma}(t_0)$, from Eq.~(\ref{Deltadef}) it is easy to see  that $\hat{\Delta}(t_0)$ is a periodic function of $t_0$ of period $2 \pi$ and, consequently, remains bounded in the limit $t_0\rightarrow +\infty$. Thus, in order to remove the secular term appearing in Eq.~(\ref{resultado3}) one has to impose the condition that  $\hat{\mathcal{F}}^{\prime}(\mathcal{T}_1)=0$ or, taking into account Eqs.~(\ref{resultado1}) and (\ref{defF}), that
\begin{eqnarray}
\frac{\partial\hat{\mathcal{G}}_{1}^{(1)}(\mathcal{T}_1)}{\partial t_1}\!\!\!\!&=&\!\!\!\!-\,i\,\hat{h}_{\mathrm{eff}}\,\hat{\mathcal{G}}_{1}^{(1)}(\mathcal{T}_1)-e^{-i t_1 \hat{h}_{\mathrm{eff}}}\frac{\partial \hat{\mathcal{G}}_{2}^{(0)}(\mathcal{T}_2)}{\partial t_2 }\nonumber\\
\!\!\!\!&&\!\!\!\!+\,\hat{\Pi}_{\mathrm{eff}}\,e^{-i t_1 \hat{h}_{\mathrm{eff}}}\hat{\mathcal{G}}_{2}^{(0)}(\mathcal{T}_2).
\end{eqnarray} 
The formal solution of the above equation is
\begin{eqnarray}
\label{resultado4}
\hat{\mathcal{G}}_{1}^{(1)}(\mathcal{T}_1)&=& e^{-i t_1 \hat{h}_{\mathrm{eff}}}\Bigg[\hat{\mathcal{G}}_{2}^{(1)}(\mathcal{T}_2)-t_1 \frac{\partial \hat{\mathcal{G}}_{2}^{(0)}(\mathcal{T}_2)}{\partial t_2 }\nonumber\\
&&+\, \hat{\Lambda}(t_1) \,\hat{\mathcal{G}}_{2}^{(0)}(\mathcal{T}_2)\Bigg],
\end{eqnarray}
where
\begin{equation}
\label{Lambdadef}
\hat{\Lambda}(t_1)=\int_{0}^{t_1} \mathrm{d}t_1^{\prime}\,e^{i t_1^{\prime} \hat{h}_{\mathrm{eff}}}\hat{\Pi}_{\mathrm{eff}}e^{-i t_1^{\prime} \hat{h}_{\mathrm{eff}}}
\end{equation}
and $\hat{\mathcal{G}}_{2}^{(1)}(\mathcal{T}_2)$ is an arbitrary function of all the time scales except $t_0$ and $t_1$. By choosing $\hat{\mathcal{G}}_{1}^{(1)}(\mathcal{T}_1)$ of the form given by Eq.~(\ref{resultado4}), we achieve to remove the secular term appearing in Eq.~(\ref{resultado3}), so that it reduces to
\begin{eqnarray}
\label{resultado3new}
\hat{\mathcal{G}}^{(2)}(\mathcal{T}_0)&=&\hat{\mathcal{G}}^{(2)}_{1}(\mathcal{T}_1)+\,\hat{\Delta}(t_0)\,\hat{\mathcal{G}}_{1}^{(0)}(\mathcal{T}_1)\nonumber\\
&& -i\,\hat{\Sigma}(t_0)\,\hat{\mathcal{G}}_{1}^{(1)}(\mathcal{T}_1),
\end{eqnarray}

In order to analyze the behavior of the operator $\hat{\Lambda}(t_1)$ in the limit $t_1\rightarrow +\infty$, it is necessary first to evaluate the operator $\hat{\Pi}_{\mathrm{eff}}$. From Eqs.~(\ref{Pidef}) and (\ref{Pieffdef}), together with Eqs.~(\ref{THamiltonian}), (\ref{effHamiltonian}) and (\ref{sigmadef}), we obtain after some lengthy calculations that
\begin{eqnarray}
\label{Pieff}
\hat{\Pi}_{\mathrm{eff}}&=&-\frac{1}{4\pi}\int_{0}^{2 \pi}\mathrm{d}t_0\int_{0}^{t_0}\mathrm{d}t_0^{\prime}\,\left[\hat{h}(t_0),\hat{h}(t_0^{\prime})\right]\nonumber \\&=&\boldsymbol{\Pi}_{\mathrm{eff}}\cdot\hat{\boldsymbol{\sigma}},
\end{eqnarray}
where $\hat{\boldsymbol{\sigma}}=\hat{\sigma}_x \mathbf{e _ x}+\hat{\sigma}_y \mathbf{e _ y}+\hat{\sigma}_z \mathbf{e _ z}$ and $\boldsymbol{\Pi}_{\mathrm{eff}}$ is the 3-D vector
\begin{eqnarray}
\boldsymbol{\Pi}_{\mathrm{eff}}&=&-\,\frac{i\, \omega_{\perp}}{4}\left\{(1+\omega_{\parallel})\left[a(r,\varphi) \mathbf{e _ x} -b(r,\varphi)\mathbf{e _ y}\right]\right.\nonumber\\ &&\left. \!-\,\omega_{\perp}  J_{0}(r)a(r,\varphi)\mathbf{e _ z} \right\}.
\end{eqnarray}
The functions $a(r,\varphi)$ and $b(r,\varphi)$ appearing in the above expression can be written in the form
 \begin{equation}
a(r,\varphi)=2 \int_{0}^{\varphi} \mathrm{d}t_0\,\sin[r \sin(t_0)]-\pi \pmb{H}_0(r)
\end{equation}
and
\begin{equation}
b(r,\varphi)=2\left[\int_{0}^{\varphi} \mathrm{d}t_0\,\cos[r \sin(t_0)]-\varphi J_0(r)\right],
\end{equation}
with $\pmb{H}_0(r)$ being the zeroth-order Struve function \cite{Abramowitz}. 

Once we have calculated $\hat{\Pi}_{\mathrm{eff}}$, the operator $\hat{\Lambda}(t_1)$ can be obtained by using the following general procedure. Let us introduce the effective frequency
\begin{equation}
\label{efffrequency}
\Omega_{\mathrm{eff}}=\sqrt{(1+\omega_{\parallel})^{2}+\left[\omega_{\perp}J_0(r)\right]^{2}},
\end{equation}
and assume for the time being that $\Omega_{\mathrm{eff}}\neq 0$. Then the effective Hamiltonian $\hat{h}_{\mathrm{eff}}$ can be expressed as
\begin{equation}
\hat{h}_{\mathrm{eff}}=-\,\frac{\Omega_{\mathrm{eff}}}{2}\, \mathbf{n}\cdot\hat{\boldsymbol{\sigma}},
\end{equation}
where $\mathbf{n}$ is the 3-D unit vector
\begin{equation}
\mathbf{n}=\Omega_{\mathrm{eff}}^{-1}\left[\omega_{\perp}J_0(r)\mathbf{e _ x}+(1+\omega_{\parallel})\mathbf{e _ z}\right].
\end{equation}

If we consider now an operator of the form $\mathbf{a}\cdot\hat{\boldsymbol{\sigma}}$, where $\mathbf{a}$ is an arbitrary 3-D vector, then it is a well-known result \cite{Galindo} that 
\begin{equation}
\label{rotation1}
e^{i t_1 \hat{h}_{\mathrm{eff}}}\mathbf{a}\cdot\hat{\boldsymbol{\sigma}}e^{-i t_1 \hat{h}_{\mathrm{eff}}}=\mathbf{a}_{\mathrm{R}}(t_1)\cdot\hat{\boldsymbol{\sigma}},
\end{equation}
where
\begin{eqnarray}
\label{rotation2}
 \mathbf{a}_{\mathrm{R}}(t_1)&=&(\mathbf{n}\cdot\mathbf{a})\mathbf{n}+\cos(\Omega_{\mathrm{eff}}t_1)\left[\mathbf{a}-
 (\mathbf{n}\cdot\mathbf{a})\mathbf{n}\right] \nonumber\\&&+\sin(\Omega_{\mathrm{eff}}t_1)\,\mathbf{n}\wedge \mathbf{a}
\end{eqnarray}
is the 3-D vector generated from $\mathbf{a}$ by a counterclockwise rotation of an angle $\Omega_{\mathrm{eff}}t_1$ about the unit vector $\mathbf{n}$, and $\wedge$ denotes the cross product of two 3-D vectors. According to Eqs.~(\ref{rotation1}) and (\ref{rotation2}), we obtain that
\begin{equation}
\label{generalresult}
\int_{0}^{t_1} \mathrm{d} t_1^{\prime}\,e^{i t_1^{\prime} \hat{h}_{\mathrm{eff}}}\mathbf{a}\cdot\hat{\boldsymbol{\sigma}}e^{-i t_1^{\prime} \hat{h}_{\mathrm{eff}}}=\mathbf{a}_{1}(t_1)\cdot\hat{\boldsymbol{\sigma}},
\end{equation}
with
\begin{eqnarray}
\mathbf{a}_1(t_1)\!\!\!\!&=&\!\!\!\!(\mathbf{n}\cdot\mathbf{a})\mathbf{n}\,t_1+ \frac{\sin(\Omega_{\mathrm{eff}}\,t_1)}{\Omega_{\mathrm{eff}}}\left[\mathbf{a}-
 (\mathbf{n}\cdot\mathbf{a})\mathbf{n}\right]\nonumber\\
 &&+\,\frac{1-\cos(\Omega_{\mathrm{eff}}\,t_1)}{\Omega_{\mathrm{eff}}}\,\mathbf{n}\wedge \mathbf{a}\,.
\end{eqnarray}
Applying the above results to the case $\mathbf{a}= \boldsymbol{\Pi}_{\mathrm{eff}}$, and taking into account that the 3-D vectors $\mathbf{n}$ and $\boldsymbol{\Pi}_{\mathrm{eff}}$ are perpendicular, we obtain that
\begin{eqnarray}
\label{resLambda}
\hat{\Lambda}(t_1)&=&\left[ \frac{\sin(\Omega_{\mathrm{eff}}\,t_1)}{\Omega_{\mathrm{eff}}}\,\boldsymbol{\Pi}_{\mathrm{eff}}\right.\nonumber\\
&&\left.+\,\frac{1-\cos(\Omega_{\mathrm{eff}}\,t_1)}{\Omega_{\mathrm{eff}}}\,\mathbf{n}\wedge \boldsymbol{\Pi}_{\mathrm{eff}}\,\right]\cdot\hat{\boldsymbol{\sigma}}.\nonumber\\
&&
\end{eqnarray}

From Eq.~(\ref{resLambda}) it is clear that, for $\Omega_{\mathrm{eff}}\neq 0$, $\hat{\Lambda}(t_1)$ is a periodic function of $t_1$ of period $\Omega_{\mathrm{eff}}$ and, consequently, remains bounded in the limit $t_1\rightarrow +\infty$. As just mentioned, this is a consequence of the fact that the 3-D vectors $\mathbf{n}$ and $\boldsymbol{\Pi}_{\mathrm{eff}}$ are perpendicular. In the case $\Omega_{\mathrm{eff}}= 0$, [i.e., if $\omega_{\parallel}=-1$ and $J_0(r)=0$ \cite{nota1}], both operators $\hat{h}_{\mathrm{eff}}$ and $\hat{\Pi}_{\mathrm{eff}}$ vanish identically and, therefore, so does $\hat{\Lambda}(t_1)$. Thus, for any value of $\Omega_{\mathrm{eff}}$ the operator $\hat{\Lambda}(t_1)$ remains bounded in the limit $t_1\rightarrow +\infty$ and, in order to remove the secular term appearing in Eq.~(\ref{resultado4}), one has to impose the condition that
\begin{equation}
\frac{\partial \hat{\mathcal{G}}_{2}^{(0)}(\mathcal{T}_2)}{\partial t_2 }=0,
\end{equation}
or, equivalently, that
\begin{equation}
\label{G02}
\hat{\mathcal{G}}_{2}^{(0)}(\mathcal{T}_2)=\hat{\mathcal{G}}_{3}^{(0)}(\mathcal{T}_3),
\end{equation} 
where $\hat{\mathcal{G}}_{3}^{(0)}(\mathcal{T}_3)$ is an arbitrary function of all the time scales except $t_0$, $t_1$ and $t_2$. With this choice of $\hat{\mathcal{G}}_{2}^{(0)}(\mathcal{T}_2)$, the secular term appearing in Eq.~(\ref{resultado4}) vanishes, so that $\hat{\mathcal{G}}_{1}^{(1)}(\mathcal{T}_1)$ is given by
\begin{eqnarray}
\label{G1new}
\hat{\mathcal{G}}^{(1)}_1(\mathcal{T}_1)\!\!\!\!&=&\!\!\!\!e^{-i t_1 \hat{h}_{\mathrm{eff}}}\left[\hat{\mathcal{G}}_{2}^{(1)}(\mathcal{T}_2)+\hat{\Lambda}(t_1)\, \hat{\mathcal{G}}_{3}^{(0)}(\mathcal{T}_3)\right].\nonumber\\
&&
\end{eqnarray}

Replacing Eq.~(\ref{G02}) in Eq.~(\ref{resultado1}) and using Eq.~(\ref{not0}), we obtain that
\begin{equation}
\label{G0}
\hat{\mathcal{G}}^{(0)}(\mathcal{T}_0)=\hat{\mathcal{G}}_{1}^{(0)}(\mathcal{T}_1)=e^{-i t_1 \hat{h}_{\mathrm{eff}}}\hat{\mathcal{G}}_{3}^{(0)}(\mathcal{T}_3).
\end{equation}
From the above expression and by a similar reasoning as the one appearing below Eq.~(\ref{resultado2}), one can see that the approximate solution (\ref{solution1}), which was obtained under the assumption that $t\ll \epsilon^{-1}$, remains valid in the wider range of time scales defined by the condition $t\ll \epsilon^{-2}$. 

\subsection{Time evolution operator for time scales $t \ll \epsilon^{-3}$}

Setting $n=3$ in Eq.~(\ref{hierarchyn}) and using Eqs.~(\ref{resultado2}), 
(\ref{Lambdadef}), (\ref{resultado3new}), (\ref{G1new}) and (\ref{G0}), one obtains that
\begin{eqnarray}
\label{order31}
i \,\frac{\partial \hat{\mathcal{G}}^{(3)}(\mathcal{T}_0)}{\partial t_0}\!\!\!\!&=&\!\!\!\!-\,i\,\hat{\mathcal{A}}(\mathcal{T}_1)+\hat{h}(t_0)\hat{\mathcal{G}}_{1}^{(2)}(\mathcal{T}_1)\nonumber\\\!\!\!\!&&\!\!\!\!+ \,i\,\hat{\Pi}(t_0)\hat{\mathcal{G}}_{1}^{(1)}(\mathcal{T}_1)-\,\hat{\Gamma}(t_0)\hat{\mathcal{G}}_{1}^{(0)}(\mathcal{T}_1),\nonumber\\
\!\!\!\!&&\!\!\!\!
\end{eqnarray}
where
\begin{equation}
\label{defA}
\hat{\mathcal{A}}(\mathcal{T}_1)=\sum_{k=1}^{3}\frac{\partial \hat{\mathcal{G}}_{1}^{(3-k)}(\mathcal{T}_1)}{\partial t_{k}}
\end{equation}
and
\begin{equation}
\label{defGamma}
\hat{\Gamma}(t_0)=\hat{\Delta}(t_0)\hat{h}_{\mathrm{eff}}-\hat{h}(t_0)\hat{\Delta}(t_0)+\hat{\Sigma}(t_0)\hat{\Pi}_{\mathrm{eff}}.
\end{equation}
The integration of Eq.~(\ref{order31}) yields
\begin{eqnarray}
\label{G31}
\hat{\mathcal{G}}^{(3)}(\mathcal{T}_0)\!\!\!\!&=& \!\!\!\!\hat{\mathcal{G}}_{1}^{(3)}(\mathcal{T}_1)-i\int_{0}^{t_0}\mathrm{d} t_0^{\prime}\left[\hat{h}(t_0^{\prime})\hat{\mathcal{G}}_{1}^{(2)}(\mathcal{T}_1)\right. \nonumber\\
\!\!\!\!&&\!\!\!\!\left.+\,i\,\hat{\Pi}(t_0^{\prime})\hat{\mathcal{G}}_{1}^{(1)}(\mathcal{T}_1)-\,\hat{\Gamma}(t_0^{\prime})\hat{\mathcal{G}}_{1}^{(0)}(\mathcal{T}_1)\right]\nonumber \\
\!\!\!\!&&\!\!\!\!-t_0\,\hat{\mathcal{A}}(\mathcal{T}_1),
\end{eqnarray}
where $\hat{\mathcal{G}}_{1}^{(3)}(\mathcal{T}_1)$ is an arbitrary function of all the time scales except $t_0$.

Taking into account that the operators $\hat{h}(t_0)$, $\hat{\Delta}(t_0)$ and $\hat{\Sigma}(t_0)$ are all periodic functions of $t_0$ of period $2\pi$, so will be the operator $\hat{\Gamma}(t_0)$. Thus, if we define the operators
\begin{equation}
\label{Gammaeffdef}
\hat{\Gamma}_{\mathrm{eff}}=\frac{1}{2 \pi} \int_{0}^{2 \pi} \mathrm{d}t_0\,\hat{\Gamma}(t_0)
\end{equation}
and
\begin{equation}
\label{Upsilondef}
\hat{\Upsilon}(t_0)=\int_{0}^{t_0} \mathrm{d}t_0^{\prime}\,\left[\hat{\Gamma}(t_0^{\prime})-\hat{\Gamma}_{\mathrm{eff}}\right],
\end{equation}
we can assure that $\hat{\Upsilon}(t_0)$ is also a periodic function of $t_0$ of period $2\pi$. Replacing the definitions (\ref{sigmadef}), (\ref{Deltadef}) and (\ref{Upsilondef}) in Eq.~(\ref{G31}), one obtains
that
\begin{eqnarray}
\label{G32}
\hat{\mathcal{G}}^{(3)}(\mathcal{T}_0)\!\!\!\!&=&\!\!\!\!\hat{\mathcal{G}}_{1}^{(3)}(\mathcal{T}_1)-t_0\,\hat{\mathcal{A}}^{\prime}(\mathcal{T}_1)-i\,\hat{\Sigma}(t_0)\hat{\mathcal{G}}_{1}^{(2)}(\mathcal{T}_1)\nonumber\\
\!\!\!\!&&\!\!\!\!+\, \hat{\Delta}(t_0)\hat{\mathcal{G}}_{1}^{(1)}(\mathcal{T}_1)+i\,\hat{\Upsilon}(t_0)\hat{\mathcal{G}}_{1}^{(0)}(\mathcal{T}_1),\nonumber\\
\!\!\!\!&&\!\!\!\!
\end{eqnarray}
with
\begin{eqnarray}
\label{defAprima}
\hat{\mathcal{A}}^{\prime}(\mathcal{T}_1)\!\!&=&\!\!\hat{\mathcal{A}}(\mathcal{T}_1)+i\,\hat{h}_{\mathrm{eff}}\,\hat{\mathcal{G}}_{1}^{(2)}(\mathcal{T}_1)\nonumber\\
\!\!&&\!\!-\,\hat{\Pi}_{\mathrm{eff}}\,\hat{\mathcal{G}}_{1}^{(1)}(\mathcal{T}_1)-i\,\,\hat{\Gamma}_{\mathrm{eff}}\,\hat{\mathcal{G}}_{1}^{(0)}(\mathcal{T}_1).
\end{eqnarray}

Since the operators $\hat{\Sigma}(t_0)$, $\hat{\Delta}(t_0)$ and $\hat{\Upsilon}(t_0)$ in Eq.~(\ref{G32}) are periodic functions of $t_0$ of period $2\pi$, they remain bounded in the limit $t_0\rightarrow +\infty$. Therefore, in order to remove the secular term appearing in Eq.~(\ref{G32}) one has to impose the condition that  $\hat{\mathcal{A}}^{\prime}(\mathcal{T}_1)=0$. Taking into account Eqs.~(\ref{Lambdadef}), (\ref{G1new}), (\ref{G0}), (\ref{defA}) and (\ref{defAprima}), the condition $\hat{\mathcal{A}}^{\prime}(\mathcal{T}_1)=0$ leads us to the differential equation
\begin{eqnarray}
\label{eqG21}
\frac{\partial \hat{\mathcal{G}}_{1}^{(2)}(\mathcal{T}_1)}{\partial t_{1}}\!\!\!\!\!&=&\!\!\!\!\!-\,i\,\hat{h}_{\mathrm{eff}}\,\hat{\mathcal{G}}_{1}^{(2)}(\mathcal{T}_1)-e^{-i t_1 \hat{h}_{\mathrm{eff}}}\Bigg[\hat{\mathcal{D}}(\mathcal{T}_2) \nonumber\\
\!\!\!\!\!&&\!\!\!\!\!\left.-\frac{\partial \hat{\Lambda}(t_1)}{\partial t_1}\,\hat{\mathcal{G}}_{2}^{(1)}(\mathcal{T}_2)-\hat{\Xi}(t_1)\hat{\mathcal{G}}_{3}^{(0)}(\mathcal{T}_3)\right],\nonumber\\
\!\!\!\!\!&&\!\!\!\!\!
\end{eqnarray}
where
\begin{equation}
\hat{\mathcal{D}}(\mathcal{T}_2)=\frac{\partial \hat{\mathcal{G}}_{2}^{(1)}(\mathcal{T}_2)}{\partial t_{2}}+\frac{\partial \hat{\mathcal{G}}_{3}^{(0)}(\mathcal{T}_3)}{\partial t_{3}}
\end{equation}
and
\begin{equation}
\label{Xidef}
\hat{\Xi}(t_1)=\frac{\partial \hat{\Lambda}(t_1)}{\partial t_1}\,\hat{\Lambda}(t_1)+i\,e^{i t_1 \hat{h}_{\mathrm{eff}}}\hat{\Gamma}_{\mathrm{eff}}\,e^{-i t_1 \hat{h}_{\mathrm{eff}}}.
\end{equation}

The formal solution of Eq.~(\ref{eqG21}) is
\begin{eqnarray}
\label{resG12first}
\hat{\mathcal{G}}_{1}^{(2)}(\mathcal{T}_1)\!\!\!\!&=&\!\!\!\!e^{-i t_1 \hat{h}_{\mathrm{eff}}}\left[\hat{\mathcal{G}}_{2}^{(2)}(\mathcal{T}_2)-t_1\,\hat{\mathcal{D}}(\mathcal{T}_2)\right.\nonumber\\
\!\!\!\!&& \!\!\!\!\left.+\,\hat{\Lambda}(t_1)\,\hat{\mathcal{G}}_{2}^{(1)}(\mathcal{T}_2)\,+\,\hat{\Theta}(t_1)\,\hat{\mathcal{G}}_{3}^{(0)}(\mathcal{T}_3)\right],\nonumber\\
\!\!\!\!&&\!\!\!\!
\end{eqnarray}
where $\hat{\mathcal{G}}_{2}^{(2)}(\mathcal{T}_2)$ is an arbitrary function of all the time scales except $t_0$ and $t_1$, and
\begin{equation}
\label{Thetadef}
\hat{\Theta}(t_1)=\int_{0}^{t_1} \mathrm{d}t_1^{\prime}\,\hat{\Xi}(t_{1}^{\prime}).
\end{equation}

In order to know the behavior of the operator $\hat{\Theta}(t_1)$ in the limit $t_1\rightarrow +\infty$, we first have to evaluate the operator $\hat{\Gamma}_{\mathrm{eff}}$. Using Eq.~(\ref{defGamma}), together with Eq.~(\ref{sigmadef}), it is easy to show that
\begin{equation}
\label{Gammaeffcalc1}
\hat{\Gamma}_{\mathrm{eff}}=\left[\hat{\Delta}_{\mathrm{eff}},\hat{h}_{\mathrm{eff}}\right]+\frac{1}{2 \pi}\int_{0}^{2 \pi} \mathrm{d}t_0\,\hat{\Sigma}(t_{0})\hat{\Pi}(t_{0}),
\end{equation}
with
\begin{equation}
\hat{\Delta}_{\mathrm{eff}}=\frac{1}{2 \pi}\int_{0}^{2 \pi} \mathrm{d}t_0\,\hat{\Delta}(t_{0}).
\end{equation}
Replacing in Eq.~(\ref{Gammaeffcalc1}) the definitions (\ref{sigmadef}) and (\ref{Pidef}), and
using Eqs.~(\ref{THamiltonian}) and (\ref{effHamiltonian}), one obtains after some lengthy calculations that
\begin{equation}
\label{resGammaeff}
\hat{\Gamma}_{\mathrm{eff}}=\left[\hat{\Delta}_{\mathrm{eff}},\hat{h}_{\mathrm{eff}}\right]+\mathbf{q}\cdot\hat{\boldsymbol{\sigma}},
\end{equation}
where
\begin{eqnarray}
\label{defq}
\mathbf{q}&=&-\left(\frac{\omega_{\perp}}{2}\right)^2\left\{\frac{\omega_{\perp}}{2} \left[\alpha_x(r,\varphi)\mathbf{e _ x}+\alpha_y(r,\varphi)\mathbf{e _ y}\right]\right.\nonumber\\
&&+\,(1+\omega_{\parallel})\alpha_z(r,\varphi)\mathbf{e _ z}\Big\},
\end{eqnarray}
with 
\begin{equation}
\label{defalphax}
\alpha_x(r,\varphi)=\frac{J_0(r)}{2}\left[a(r,\varphi)\right]^2-\gamma_1(r),
\end{equation}
\begin{equation}
\label{defalphay}
\alpha_y(r,\varphi)=-\,\frac{J_0(r)}{2}\,a(r,\varphi)\,b(r,\varphi)
\end{equation}
and
\begin{equation}
\label{defalphaz}
\alpha_z(r,\varphi)=\left[\frac{a(r,\varphi)}{2}\right]^2+\left[\frac{b(r,\varphi)}{2}\right]^2-\gamma_2(r).
\end{equation}
In the above expressions $\gamma_1(r)$ and $\gamma_2(r)$ are two functions of the parameter $r$ which are given by
\begin{eqnarray}
\label{defgamma1}
\gamma_1(r)\!\!\!\!&=&\!\!\!\!\frac{\pi^2}{2}\, J_0(r)\left[\pmb{H}_0(r)\right]^2\nonumber\\
\!\!\!\!&&\!\!\!\!+\,\frac{2}{\pi}\int_{0}^{2 \pi}\mathrm{d}\varphi\int_{0}^{\varphi}\mathrm{d}\varphi^{\prime}
\int_{0}^{\varphi}\mathrm{d}\varphi^{\prime\prime}\sin(r \sin \varphi)\nonumber\\
\!\!\!\!&&\!\!\!\!\times \cos(r \sin \varphi^{\prime}) \sin(r \sin \varphi^{\prime\prime})\, 
\end{eqnarray}
and
\begin{eqnarray}
\label{defgamma2}
\gamma_2(r)\!\!\!\!&=&\!\!\!\!\frac{\pi^2}{4} \left[\pmb{H}_0(r)\right]^2-\frac{4\pi^2}{3}\left[J_0(r)\right]^2\nonumber\\
\!\!\!\!&&\!\!\!\!-\,\frac{J_0(r)}{2 \pi}\,\int_{0}^{2 \pi}\!\!\!\!\mathrm{d}\varphi \,\varphi^2\,\cos(r \sin \varphi)\nonumber\\
\!\!\!\!&&\!\!\!\!+\frac{1}{\pi}\int_{0}^{2 \pi}\!\!\!\!\mathrm{d}\varphi\int_{0}^{\varphi}\!\!\!\!\mathrm{d}\varphi^{\prime}\,\varphi\, \cos[r(\sin \varphi-\sin \varphi^{\prime})].\nonumber\\
\!\!\!\!&&\!\!\!\!
\end{eqnarray}

As in the previous subsection, now we will consider separately the cases $\Omega_{\mathrm{eff}}\neq 0$ and $\Omega_{\mathrm{eff}}= 0$. For $\Omega_{\mathrm{eff}}\neq 0$, the first term of the right-hand side of Eq.~(\ref{Xidef}) can be easily evaluated using Eq.~(\ref{resLambda}). The second term can be calculated from Eq.~(\ref{resGammaeff}), making use of Eqs.~(\ref{rotation1}) and (\ref{rotation2}) with $\mathbf{a}=\mathbf{q}$. Replacing the results so obtained in Eq.~(\ref{Thetadef}), one finally finds that
\begin{equation}
\label{resTheta}
\hat{\Theta}(t_1)=-\,i\,\eta \, t_1 \,\hat{h}_{\mathrm{eff}}+\hat{\Phi}(t_1),
\end{equation}
where
\begin{eqnarray}
\label{etadef}
\eta \!\!\!\!&=&\!\!\!\!\frac{2}{\Omega_{\mathrm{eff}}}\left(\mathbf{n}\cdot\mathbf{q}-\frac{\boldsymbol{\Pi}_{\mathrm{eff}}\cdot
\boldsymbol{\Pi}_{\mathrm{eff}}}{\Omega_{\mathrm{eff}}}
\right)\nonumber\\
\!\!\!\!&=&\!\!\!\!\frac{1}{2}\left(\frac{\omega_{\perp}}{\Omega_{\mathrm{eff}}}\right)^2\!\!\left[\frac{\omega_{\perp}^2}{2} J_0(r) \gamma_1(r)+(1+\omega_{\parallel})^2\gamma_2(r)\right]\nonumber\\
&&
\end{eqnarray}
and
\begin{eqnarray}
\label{Phidef}
\hat{\Phi}(t_1)\!\!\!\!&=&\!\!\!\!i\,\int_{0}^{t_1} \mathrm{d}t_1^{\prime}\,e^{i t_1^{\prime} \hat{h}_{\mathrm{eff}}}\left[\hat{\Delta}_{\mathrm{eff}},\hat{h}_{\mathrm{eff}}\right]\,e^{-i t_1^{\prime} \hat{h}_{\mathrm{eff}}}\nonumber\\\!\!\!\!&&\!\!\!\!+\,i\, \frac{\sin(\Omega_{\mathrm{eff}}\,t_1)}{\Omega_{\mathrm{eff}}}\left[\mathbf{q}\cdot\hat{\boldsymbol{\sigma}}-\eta\, \hat{h}_{\mathrm{eff}}\right]\nonumber\\
\!\!\!\! &&\!\!\!\!+\,\frac{1-\cos(\Omega_{\mathrm{eff}}\,t_1)}{\Omega_{\mathrm{eff}}}\,\left\{\left[\mathbf{n}\cdot\mathbf{q}-\frac{\eta \,\Omega_{\mathrm{eff}}}{2} \right]\hat{I}\right.\nonumber\\ 
\!\!\!\! &&\!\!\!\!+\,i\,(\mathbf{n}\wedge \mathbf{q})\cdot\hat{\boldsymbol{\sigma}} \bigg\}
.
\end{eqnarray}
Notice that the first term of the right-hand side of Eq.~(\ref{Phidef}) is a periodic function of $t_1$ of period $\Omega_{\mathrm{eff}}$. In order to see that this is so, one just has to apply the same reasoning as for the operator $\hat{\Lambda}(t_1)$, taking into account that, as can be easily shown,  $[\hat{\Delta}_{\mathrm{eff}},\hat{h}_{\mathrm{eff}}]=i\,\mathbf{p} \cdot\hat{\boldsymbol{\sigma}}$ with 
$\mathbf{p}$ being a certain 3D-vector perpendicular to  $ \mathbf{n}$. According to Eq.~(\ref{Phidef}), we can therefore assert that $\hat{\Phi}(t_1)$ is a periodic function of $t_1$ of period $\Omega_{\mathrm{eff}}$. 

Replacing Eq.~(\ref{resTheta}) in Eq.~(\ref{resG12first}), it results 
\begin{eqnarray}
\label{resG12second}
\hat{\mathcal{G}}_{1}^{(2)}(\mathcal{T}_1)\!\!\!\!&=&\!\!\!\!e^{-i t_1 \hat{h}_{\mathrm{eff}}}\left[\hat{\mathcal{G}}_{2}^{(2)}(\mathcal{T}_2)-t_1\,
\hat{\mathcal{D}}^{\prime}(\mathcal{T}_2)\right.\nonumber\\
\!\!\!\!&& \!\!\!\!\left.+\,\hat{\Lambda}(t_1)\,\hat{\mathcal{G}}_{2}^{(1)}(\mathcal{T}_2)\,+\,\hat{\Phi}(t_1)\,\hat{\mathcal{G}}_{3}^{(0)}(\mathcal{T}_3)\right],\nonumber\\
\!\!\!\!&&\!\!\!\!
\end{eqnarray}
where
\begin{equation}
\hat{\mathcal{D}}^{\prime}(\mathcal{T}_2)=\hat{\mathcal{D}}(\mathcal{T}_2)+i\,\eta \,\hat{h}_{\mathrm{eff}}\hat{\mathcal{G}}_{3}^{(0)}(\mathcal{T}_3).
\end{equation}
Since both $\hat{\Lambda}(t_1)$ and $\hat{\Phi}(t_1)$ are periodic functions of $t_1$ of period $\Omega_{\mathrm{eff}}$, they remain bounded in the limit $t_1 \rightarrow +\infty$. Therefore, in order to remove the secular term appearing in Eq.~(\ref{resG12second}) one has to impose the condition that  $\hat{\mathcal{D}}^{\prime}(\mathcal{T}_2)=0$ or, equivalently, that
\begin{eqnarray}
\frac{\partial \hat{\mathcal{G}}_{2}^{(1)}(\mathcal{T}_2)}{\partial t_{2}}\!\!\!&=&\!\!\!\,-\,\left[\frac{\partial \hat{\mathcal{G}}_{3}^{(0)}(\mathcal{T}_3)}{\partial t_{3}}+i\,\eta \,\hat{h}_{\mathrm{eff}}\hat{\mathcal{G}}_{3}^{(0)}(\mathcal{T}_3)\right].\nonumber\\
&&
\end{eqnarray}
The integration of the above equation leads to
\begin{eqnarray}
\label{resG21}
\hat{\mathcal{G}}_{2}^{(1)}(\mathcal{T}_2)\!\!\!\!&=&\!\!\!\! \hat{\mathcal{G}}_{3}^{(1)}(\mathcal{T}_3)\nonumber\\
\!\!\!\!&&\!\!\!\! -\,t_2 \left[\frac{\partial \hat{\mathcal{G}}_{3}^{(0)}(\mathcal{T}_3)}{\partial t_{3}}+i\,\eta \,\hat{h}_{\mathrm{eff}}\hat{\mathcal{G}}_{3}^{(0)}(\mathcal{T}_3)\right],\nonumber\\
\!\!\!\!&&\!\!\!\!
\end{eqnarray}
where $\hat{\mathcal{G}}_{3}^{(1)}(\mathcal{T}_3)$ is an arbitrary function of all the time scales except $t_0$, $t_1$ and $t_2$. 

To avoid the secular term appearing in Eq.~(\ref{resG21}) we have to assume that
\begin{equation}
\frac{\partial \hat{\mathcal{G}}_{3}^{(0)}(\mathcal{T}_3)}{\partial t_{3}}=-\,i\,\eta \,\hat{h}_{\mathrm{eff}}\hat{\mathcal{G}}_{3}^{(0)}(\mathcal{T}_3),
\end{equation}
whose formal solution is
\begin{equation}
\hat{\mathcal{G}}_{3}^{(0)}(\mathcal{T}_3)=e^{-i \eta t_3  \hat{h}_{\mathrm{eff}}} \hat{\mathcal{G}}_{4}^{(0)}(\mathcal{T}_4),
\end{equation}
with $\hat{\mathcal{G}}_{4}^{(0)}(\mathcal{T}_4)$ being an arbitrary function of all the time scales except $t_0$, $t_1$, $t_2$ and $t_3$. Replacing the above result in Eq.~(\ref{G0}), we finally obtain that, for $\Omega_{\mathrm{eff}}\neq 0$,  $\hat{\mathcal{G}}^{(0)}(\mathcal{T}_0)$ is given by
\begin{equation}
\label{G0new}
\hat{\mathcal{G}}^{(0)}(\mathcal{T}_0)=e^{-i (t_1+\eta t_3 ) \hat{h}_{\mathrm{eff}}}\hat{\mathcal{G}}_{4}^{(0)}(\mathcal{T}_4).
\end{equation}

Let us consider now the case in which $r$ is a zero of the Bessel function $J_0(r)$ and $\omega_{\parallel}=-1$, so that $\Omega_{\mathrm{eff}}=0$. Specifically, we will assume that $r=r_j$, where $r_j$ is the $j$-th zero of $J_0(r)$ with $j=1,\,2,\,\dots$. As mentioned in the previous subsection, in this case the operators $\hat{h}_{\mathrm{eff}}$, $\hat{\Pi}_{\mathrm{eff}}$ and $\hat{\Lambda}(t_1)$ vanish identically. Thus, from Eqs.~(\ref{Xidef}), (\ref{Thetadef}),  (\ref{resGammaeff}), (\ref{defq}), (\ref{defalphax}) and (\ref{defalphay}) it is quite easy to see that
\begin{equation}
\label{resTheta0}
\hat{\Theta}(t_1)=\,i\,\hat{\Gamma}_{\mathrm{eff}} \,t_1=\,i\,\left(\frac{\omega_{\perp}}{2} \right)^3 \gamma_{1,j} \,t_1\,\hat{\sigma}_x\,,
\end{equation}
where $\gamma_{1,j}=\gamma_1(r_j)$, i.e.,
\begin{eqnarray}
\label{defgamma1j}
\gamma_{1,j}\!\!\!\!&=&\!\!\!\!\frac{2}{\pi}\int_{0}^{2 \pi}\mathrm{d}\varphi\int_{0}^{\varphi}\mathrm{d}\varphi^{\prime}
\int_{0}^{\varphi}\mathrm{d}\varphi^{\prime\prime}\sin(r_j \sin \varphi)\nonumber\\
\!\!\!\!&&\!\!\!\!\times \cos(r_j \sin \varphi^{\prime}) \sin(r_j \sin \varphi^{\prime\prime}).
\end{eqnarray}

Replacing Eq.~(\ref{resTheta0}) in Eq.~(\ref{resG12first}), and taking into account that $\hat{\Lambda}(t_1)=0$, one obtains that
\begin{equation}
\label{G21newnew}
\hat{\mathcal{G}}_{1}^{(2)}(\mathcal{T}_1)= e^{-i t_1 \hat{h}_{\mathrm{eff}}}\left[\hat{\mathcal{G}}_{2}^{(2)}(\mathcal{T}_2)-t_1\,
\hat{\mathcal{D}}^{\prime}(\mathcal{T}_2)\right],
\end{equation}
with
\begin{equation}
\hat{\mathcal{D}}^{\prime}(\mathcal{T}_2)=\hat{\mathcal{D}}(\mathcal{T}_2)-i\,\left(\frac{\omega_{\perp}}{2} \right)^3 \gamma_{1,j} \,\hat{\sigma}_x\,\hat{\mathcal{G}}_{3}^{(0)}(\mathcal{T}_3)\,.
\end{equation}
Then, following the same procedure as in the case $\Omega_{\mathrm{eff}}\neq 0$, it results that
\begin{eqnarray}
\label{finalres2}
\hat{\mathcal{G}}^{(0)}(\mathcal{T}_0)\!\!\!\!&=&\!\!\!\!\hat{\mathcal{G}}_{3}^{(0)}(\mathcal{T}_3)=e^{i \left(\frac{\omega_{\perp}}{2} \right)^3 \gamma_{1,j} \,\hat{\sigma}_x t_3} \hat{\mathcal{G}}_{4}^{(0)}(\mathcal{T}_4)\nonumber\\
&&
\end{eqnarray}
when $\Omega_{\mathrm{eff}}= 0$.

Now we apply the same reasoning as at the end of subsection \ref{MS1}. Let us assume that we are only interested in time values such that $t\ll \epsilon^{-3}$, so that, according to Eq.~(\ref{timescales}),  it is fulfilled that $t_k\ll 1$ for $k\geq 4$. Then, in Eqs.~(\ref{G0new}) and (\ref{finalres2}), it is possible to replace the function $\hat{\mathcal{G}}_{4}^{(0)}(\mathcal{T}_4)$ by a constant to be determined from the initial condition in Eq.~(\ref{IC2}). Expressing $t_1$ and $t_3$ in terms of $t$ and taking into account Eq.~(\ref{series}), we finally obtain that
\begin{equation}
\label{resultadofinal1}
\hat{G}(t)= e^{-i t \hat{h}_{\mathrm{ms}}}e^{-i \theta(0)\hat{\sigma}_ z/2}+\mathcal{O}(\epsilon),
\end{equation}
where $\hat{h}_{\mathrm{ms}}$ is a multiple scale Hamiltonian defined by
\begin{equation}
\label{MSHamiltoniann0}
\hat{h}_{\mathrm{ms}}=(1+\epsilon^2 \eta)\,\hat{h}_{\mathrm{eff}}
\end{equation}
in the case that $\Omega_{\mathrm{eff}}\neq 0$, and by
\begin{equation}
\label{MSHamiltonian0}
\hat{h}_{\mathrm{ms}}=-\,\epsilon^2\left(\frac{\omega_{\perp}}{2} \right)^3 \gamma_{1,j} \,\hat{\sigma}_x
\end{equation}
in the case that $\Omega_{\mathrm{eff}}= 0$ and $r=r_j$. 

Finally, according to Eq.~(\ref{UTransformation}), the time evolution operator obtained by the multiple scale method and valid for time scales $t\ll \epsilon^{-3}$ is
\begin{equation}
\label{MSsolution}
\hat{U}_{\mathrm{ms}}(t)=e^{-i \left[t -\theta(t)\right]\hat{\sigma}_ z/2}e^{-i t\hat{h}_{\mathrm{ms}}}e^{-i \theta(0)\hat{\sigma}_ z/2}.
\end{equation}
The above three expressions, together with the definitions of $\eta$ [Eq.~(\ref{etadef})] and $\gamma_{1,j}$ [Eq.~(\ref{defgamma1j})] are the main results of this section.
 
\section{Results}
\label{Results}
We begin this section with a brief description of the resonant behavior displayed by a spin-1/2 particle under the influence of the rotating magnetic field in Eq.~(\ref{MF}), when the high-frequency magnetic field  in Eq.~(\ref{HFF}) is absent (i.e., when $r=0$). Then, according to Eqs.~(\ref{defthetadet}) and (\ref{THamiltonian}), the Hamiltonian $\hat{h}(\Omega_{\mathrm{HF}}t)$ becomes time independent and equal to
\begin{equation}
\label{defh0}
\hat{h}_0=-\frac{1}{2}\left[\omega_{\perp}\hat{\sigma}_x+(1+\omega_{\parallel})\hat{\sigma}_z\right].
\end{equation}
Taking into account Eqs.~(\ref{UTransformation}) and (\ref{evolutionoperator2}), the time evolution operator will be then given by
\begin{equation}
\label{Tevolutionoperator0}
\hat{U}(t)=e^{-i t \hat{\sigma}_ z/2}e^{-i t\hat{h}_{0}}.
\end{equation}

Henceforth, we will be interested in studying the time dependence of the expected value of $\hat{\sigma}_ z$. If we take as initial condition an arbitrary state $|\Psi,0\rangle$, this expected value can easily be evaluated from the time evolution operator by using the expression 
\begin{equation}
\label{expforexpvalsigz}
\langle \hat{\sigma}_ z;t\rangle= \langle \Psi,0| \hat{U}^{\dagger}(t)\hat{\sigma}_ z\hat{U}(t)|\Psi,0\rangle.
\end{equation}
Let us assume that we take as initial condition $|\Psi,0\rangle=|\pm \rangle$, with $|\pm \rangle$ being the eigenvectors of $\hat{\sigma}_ z$ corresponding to the eigenvalues $\pm 1$. Then, from Eqs.~(\ref{Tevolutionoperator0}) and (\ref{expforexpvalsigz}), together with Eqs.~(\ref{rotation1}) and (\ref{rotation2}) with $\hat{h}_{\mathrm{eff}}$ replaced by $\hat{h}_{0}$ and $t_1$ by $t$, it is easy to see that
\begin{equation}
\label{expvalsigma0}
\langle \hat{\sigma}_z;t\rangle^{(\pm)}=\pm \left\{1 + \frac{\omega_{\perp}^2}{\Omega_0^2} \left[\cos(\Omega_0 t)-1\right]\right\},
\end{equation}
where $\langle \hat{\sigma}_z;t\rangle^{(\pm)}=\langle \pm| \hat{U}^{\dagger}(t)\hat{\sigma}_ z\hat{U}(t)|\pm\rangle$ and $\Omega_0=\sqrt{(1+\omega_{\parallel})^2+\omega_{\perp}^2}$. 

\begin{figure}[htb]
\vspace{0.5cm}
\includegraphics[scale=0.30]{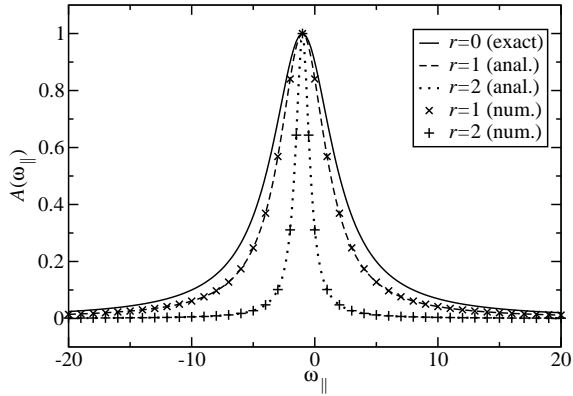}
\caption{Resonant behavior of the amplitude $A(\omega_{\parallel})$ for $\omega_{\perp}=3$ and $r=0$, $1$ and $2$. Solid line: exact result for the case $r=0$ obtained from Eq.~(\ref{amplitude0}). Dashed and dotted lines: analytical results  obtained from Eq.~(\ref{amplitudernonzero}) for $r=1$ (dashed line) and $r=2$ (dotted line). Symbols: results obtained from the numerical solution of Eqs.~(\ref{evolutionoperator1})-(\ref{Hamiltonian}), and the use of Eq.~(\ref{expforexpvalsigz}) with $|\Psi,0\rangle=|+ \rangle$.  The parameter values of the high-frequency magnetic field used in the numerical calculations are: $\Omega_{\mathrm{HF}}=50$, $\varphi=\pi/2$, $r=1$ (crosses) and $r=2$ (pluses).}
\label{resonance1}
\end{figure}

From the exact expression (\ref{expvalsigma0}) one can see that, in the absence of the high-frequency magnetic field, $\langle \hat{\sigma}_z;t\rangle^{(\pm)}$ is a periodic function of time of period  $2 \pi/\Omega_0$. The average of this function over a period is $\pm (1+\omega_{\parallel})^2/\Omega_0^2$, and the amplitude of the oscillations is given by
\begin{equation}
\label{amplitude0}
A=\frac{\omega_{\perp}^2}{\Omega_0^2}=\frac{\omega_{\perp}^2}{(1+\omega_{\parallel})^2+\omega_{\perp}^2}.
\end{equation}
This amplitude as a function of $\omega_{\parallel}$ displays a resonant behavior with a maximum at $\omega_{\parallel}=-1$. This behavior is depicted with a solid line in Fig.~\ref{resonance1} for the case $\omega_{\perp}=3$.

  
When the high-frequency magnetic field is present, we do not have an exact expression for the time evolution operator and have to resort to Eqs.~(\ref{AMsolution2}) or (\ref{MSsolution}), depending on whether one uses the averaging method or the multiple scale method. Let us consider first the case in which $r$ is not a zero of $J_0(r)$, so that $\Omega_{\mathrm{eff}}\neq 0$. Then, from Eqs.~(\ref{effHamiltonian}), (\ref{AMsolution2}) and  (\ref{expforexpvalsigz}), together with Eqs.~(\ref{rotation1}) and (\ref{rotation2}), the averaging method yields
\begin{eqnarray}
\label{expvalsigmaav}
\langle \hat{\sigma}_z;t\rangle_{\mathrm{av}}^{(\pm)}\!\!\!\!\!&=\!\!\!\!\!&\pm \left\{1 + \frac{\left[\omega_{\perp} J_0(r)\right]^2}{\Omega_{\mathrm{eff}}^2} \left[\cos(\Omega_{\mathrm{eff}} t)-1\right]\right\}.\nonumber\\
\!\!\!\!\!&&\!\!\!\!\!
\end{eqnarray}

However, if one uses the multiple scale method, from Eqs.~(\ref{MSHamiltoniann0}) and (\ref{MSsolution}) one obtains
\begin{eqnarray}
\label{expvalsigmams}
\langle \hat{\sigma}_z;t\rangle_{\mathrm{ms}}^{(\pm)}\!\!\!\!\!&=\!\!\!\!\!&\pm \left\{1 + \frac{\left[\omega_{\perp} J_0(r)\right]^2}{\Omega_{\mathrm{eff}}^2} \left[\cos(\Omega_{\mathrm{ms}} t)-1\right]\right\},\nonumber\\
\!\!\!\!\!&&\!\!\!\!\!
\end{eqnarray}
with the multiple scale frequency
\begin{equation}
\label{MSfrequencynon0}
\Omega_{\mathrm{ms}}=(1+\epsilon^2 \eta)\Omega_{\mathrm{eff}}.
\end{equation}
Notice that both methods lead to the same expression 
\begin{equation}
\label{amplitudernonzero}
A=\frac{\left[\omega_{\perp}J_0(r)\right]^2}{\Omega_{\mathrm{eff}}^2}=\frac{\left[\omega_{\perp}J_0(r)\right]^2}{(1+\omega_{\parallel})^2+\left[\omega_{\perp}J_0(r)\right]^2}
\end{equation}
for the amplitude of the oscillations of the function  $\langle \hat{\sigma}_z;t\rangle^{(\pm)}$.

According to Eq.~(\ref{amplitudernonzero}), the resonant peak at  $\omega_{\parallel}=-1$ persists in the presence of a high-frequency magnetic field when $r$ is not a zero of $J_0(r)$. Since the width of the peak is related to $\omega_{\perp}J_0(r)$, the presence of the high-frequency magnetic field gives rise to a narrowing of the resonant peak with respect to the case $r=0$. These analytical results are shown in Fig.~\ref{resonance1} for $\omega_{\perp}=3$ with a dashed line ($r=1$) and a dotted line ($r=2$). In Fig.~\ref{resonance1}, we also show the amplitude values obtained from the numerical solution of Eqs.~(\ref{evolutionoperator1})-(\ref{Hamiltonian}), and the use of Eq.~(\ref{expforexpvalsigz}) with $|\Psi,0\rangle=|+ \rangle$. The numerical integration of the differential equation (\ref{evolutionoperator1}) has been performed using the function \texttt{NDSolve} of the software package \textsc{Mathematica} with the option \texttt{MaxSteps} $\rightarrow \infty$. In order to have a well-defined amplitude value, the  high-frequency and small-amplitude oscillations observed in the numerical values of $\langle \hat{\sigma}_z;t\rangle^{(+)}$ have been removed by performing a time average of $\langle \hat{\sigma}_z;t\rangle^{(+)}$ over a high-frequency period. The parameter values  of the high-frequency magnetic field used in the numerical calculations are: $\Omega_{\mathrm{HF}}=50$, $\varphi=\pi/2$, $r=1$ (crosses) and $r=2$ (pluses). Notice that there is a very good agreement between the analytical and the numerical results.

Although the analytical expressions for $\langle \hat{\sigma}_z;t\rangle^{(\pm)}$ obtained by the averaging method [Eq.~(\ref{expvalsigmaav})] and the multiple scale method [Eq.~(\ref{expvalsigmams})] look very similar, there exists a slight difference $\Omega_{\mathrm{ms}}-\Omega_{\mathrm{eff}}=\epsilon^2 \eta \Omega_{\mathrm{eff}}$ between the frequencies of the oscillations predicted by both methods. This slight difference becomes important for large values of $t$. In particular, for time scales $t\approx \pi/|\epsilon^2 \eta \Omega_{\mathrm{eff}}|$,  the expected values $\langle \hat{\sigma}_z;t\rangle_{\mathrm{ms}}^{(\pm)}$ and $\langle \hat{\sigma}_z;t\rangle_{\mathrm{av}}^{(\pm)}$ are $\pi$ radians out of phase with each other. For instance, for $\omega_{\perp}=3$, $\omega_{\parallel}=-1$, $r=1$ and $\Omega_{\mathrm{HF}}=50$, this situation occurs for $t\approx \pi/|\epsilon^2 \eta \Omega_{\mathrm{eff}}|\approx 1700$. To work out this number, we have used the value $\gamma_1(1)\approx-0.684533$, obtained numerically from Eq.~(\ref{defgamma1}), for the calculation of $\eta$ [see Eq.~(\ref{etadef})]. The integral appearing in the definition of $\gamma_1(1)$ [see Eq.~(\ref{defgamma1})] has been evaluated numerically using the function \texttt{NIntegrate} of the computer algebra program \textsc{Mathematica}.

\begin{figure}[htb]
\vspace{0.5cm}
\includegraphics[scale=0.29]{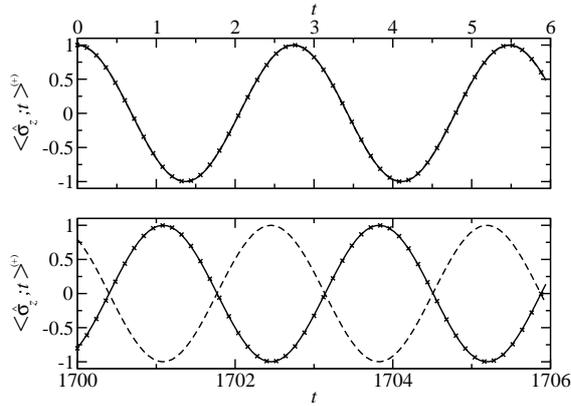}
\caption{Time dependence of $\langle \hat{\sigma}_z;t\rangle^{(+)}$ in the time intervals $t\in [0,6]$ (upper panel) and $t\in[1700,1706]$ (lower panel). Dashed lines: analytical results obtained from Eq.~(\ref{expvalsigmaav}) (averaging method). Solid lines: analytical results obtained from Eq.~(\ref{expvalsigmams}) (multiple scale method). Crosses: results obtained from the numerical solution of Eqs.~(\ref{evolutionoperator1})-(\ref{Hamiltonian}), and the use of Eq.~(\ref{expforexpvalsigz}) with $|\Psi,0\rangle=|+ \rangle$. The parameter values are: $\omega_{\perp}=3$, $\omega_{\parallel}=-1$, $r=1$, $\Omega_{\mathrm{HF}}=50$ and $\varphi=\pi/2$. In the upper panel the dashed line is hidden by the solid line.}
\label{timebehavior1}
\end{figure}

In Fig.~\ref{timebehavior1}, it is shown the time dependence of $\langle \hat{\sigma}_z;t\rangle^{(+)}$
in the intervals $t\in[0,6]$ (upper panel) and $t\in[1700, 1706]$ (lower panel), for the above mentioned parameter values. The results obtained from Eq.~(\ref{expvalsigmaav}) (averaging method) have been plotted with dashed lines, and those obtained from Eq.~(\ref{expvalsigmams}) (multiple scale method) with solid lines. We also depict with crosses the results obtained from the numerical solution of Eqs.~(\ref{evolutionoperator1})-(\ref{Hamiltonian}) (for $\varphi=\pi/2$), and the use of Eq.~(\ref{expforexpvalsigz}) with $|\Psi,0\rangle=|+ \rangle$. For short time scales (upper panel), both the averaging and the multiple scale methods agree very well with the numerical results. In fact, the dashed line is completely hidden by the solid one. Nevertheless, in the time interval $[1700, 1706]$
only the multiple scale method reproduces the numerical results, whereas the averaging method predicts a time dependence which is approximately $\pi$ radians out of phase with respect to these ones.

Now let us consider the case in which $r$ is equal to one of the zeros of the Bessel function $J_0(r)$, for instance, the $j$-th zero $r_j$. If we also assume that $\omega_{\parallel}\neq -1$, it is easy to see that Eqs.~(\ref{expvalsigmaav}) and (\ref{expvalsigmams}) remain valid, since the condition that $\Omega_{\mathrm{eff}}\neq 0$ still holds. Consequently, from Eqs.~(\ref{expvalsigmaav}) and (\ref{expvalsigmams}) one obtains that
\begin{equation}
\label{simazavmsrnonzero}
\langle \hat{\sigma}_z;t\rangle_{\mathrm{av}}^{(\pm)}=\langle \hat{\sigma}_z;t\rangle_{\mathrm{ms}}^{(\pm)}=\pm 1,
\end{equation}
for $r=r_j$ and $\omega_{\parallel}\neq -1$. 

For $r=r_j$ and $\omega_{\parallel}= -1$ the averaging and the multiple scale methods lead to different expressions for $\langle\hat{\sigma}_z;t\rangle^{(\pm)}$. If one uses the averaging method and takes into account Eqs.~(\ref{AMsolution2}) and (\ref{expforexpvalsigz}), the result is  $\langle\hat{\sigma}_z;t\rangle_{\mathrm{av}}^{(\pm)}=\pm 1$, since the effective Hamiltonian in Eq.~(\ref{effHamiltonian}) vanishes when $r=r_j$ and $\omega_{\parallel}= -1$. By contrast, if one uses the multiple scale method and takes into account that for these parameter values $\Omega_{\mathrm{eff}}= 0$, from Eqs.~(\ref{MSHamiltonian0}) and (\ref{MSsolution}) one obtains
\begin{equation}
\label{msexpvalsigzrzero}
\langle \hat{\sigma}_z;t\rangle_{\mathrm{ms}}^{(\pm)}=\pm \cos\left(\Omega_{\mathrm{ms},j} t\right),
\end{equation}
with the multiple scale frequency
\begin{equation}
\label{msfrequencyrzero}
\Omega_{\mathrm{ms},j}=\frac{\epsilon^2}{4}\,\omega_{\perp}^3 \, \gamma_{1,j}\,.
\end{equation}

\begin{figure}[htb]
\includegraphics[scale=0.3]{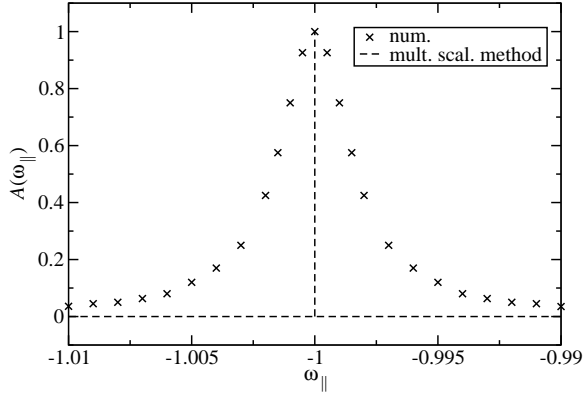}
\caption{Resonant behavior of the amplitude $A(\omega_{\parallel})$ for $r=r_1\approx 2.40483$ [first zero of the Bessel function $J_0(r)$]. Dashed line: analytical results obtained by the multiple scale method [Eq.~(\ref{mswidthzero})]. Crosses: numerical results obtained following the same procedure as in Fig.~\ref{resonance1}. The parameter values used in the numerical calculations are: $\omega_{\perp}=3$, $\Omega_{\mathrm{HF}}=50$ and $\varphi=\pi/2$. The scale of abscissa has been magnified with respect to Fig.~\ref{resonance1} in order to make visible the finite width of the resonant behavior observed numerically.}
\label{resonance2}
\end{figure}

From these results one concludes that, for $r=r_j$, the averaging method predicts that $A_{\mathrm{av}}(\omega_{\parallel})=0$ for all the values of $\omega_{\parallel}$ and, consequently, the complete disappearance of the resonant behavior. By contrast, the multiple scale method predicts that
\begin{equation}
\label{mswidthzero}
A_{\mathrm{ms}}(\omega_{\parallel})=\left\{ 
\begin{array}{l l}
0 & \quad \mbox{if \,\,$\omega_{\parallel}\neq -1$}\\
1 & \quad \mbox{if \,\,$\omega_{\parallel}= -1$},\\
\end{array}
\right.
\end{equation}
i.e., the resonant peak at $\omega_{\parallel}= -1$ persists but its width becomes zero. This last behavior is depicted in Fig.~\ref{resonance2} with a dashed line. In Fig.~\ref{resonance2}, we also show with crosses the amplitude values obtained numerically, by the numerical procedure described previously, for the parameter values $\omega_{\perp}=3$, $r=r_1\approx 2.40483$, $\Omega_{\mathrm{HF}}=50$ and $\varphi=\pi/2$. The numerical results highlight the existence of an extremely narrow peak around  $\omega_{\parallel}=-1$ and, consequently, confirm the prediction made by the multiple scale method. Nevertheless, the multiple scale method is not able to explain the finite width of the resonant peak observed in the numerical results shown in Fig.~\ref{resonance2}. This fact shows up the inability of the analytical expression in Eq.~(\ref{simazavmsrnonzero}) to describe
properly the time behavior of $\langle\hat{\sigma}_z;t\rangle^{(\pm)}$ in a small neighborhood around  $\omega_{\parallel}=-1$. This inability is not surprising if one takes into account that the expression for $\langle\hat{\sigma}_z;t\rangle_{\mathrm{av}}^{(\pm)}$ in Eq.~(\ref{simazavmsrnonzero}) does not converge to the result in Eq.~(\ref{msexpvalsigzrzero}) in the limit $\omega_{\parallel}\rightarrow -1$. In order to avoid this discontinuity, we suspect that it is necessary to extend the multiple scale method described in Sec.~\ref{MSsection} to a larger number of time scales.

\begin{figure}[htb]
\vspace{0.5cm}
\includegraphics[scale=0.29]{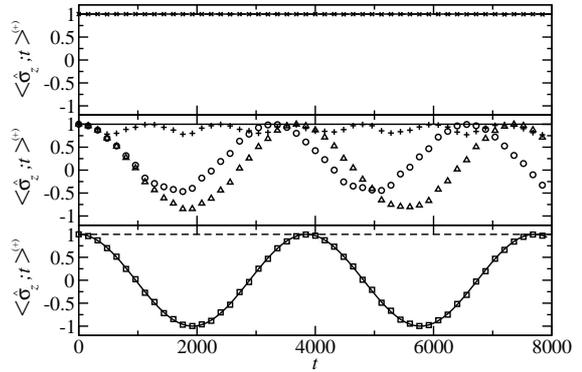}
\caption{Time dependence of $\langle \hat{\sigma}_z;t\rangle^{(+)}$ for $r=r_1\approx 2.40483$ [first zero of the Bessel function $J_0(r)$] and five values of $\omega_{\parallel}$ around $\omega_{\parallel}=-1$. The rest of the parameter values are: 
$\omega_{\perp}=3$, $\Omega_{\mathrm{HF}}=50$ and $\varphi=\pi/2$. Dashed lines: analytical results obtained from Eq.~(\ref{simazavmsrnonzero}) for all the values of $\omega_{\parallel}$ (averaging method). Solid lines: analytical results obtained from Eq.~(\ref{simazavmsrnonzero}) for $\omega_{\parallel}\neq -1$, and Eq.~(\ref{msexpvalsigzrzero}) for $\omega_{\parallel}= -1$ (multiple scale method). Symbols: numerical results. Upper panel: $\omega_{\parallel}=-1.1$ (crosses). Middle panel: $\omega_{\parallel}=-1.005$ (pluses), $\omega_{\parallel}=-1.001$ (circles) and $\omega_{\parallel}=-1.0005$ (triangles). Lower panel: $\omega_{\parallel}=-1$ (squares).  In the upper and middle panels the dashed lines are hidden by the solid lines.}
\label{timebehavior2}
\end{figure}

To illustrate this point, in Fig.~\ref{timebehavior2} it is shown the time dependence of $\langle \hat{\sigma}_z;t\rangle^{(+)}$ for $r=r_1\approx 2.40483$ and five values of $\omega_{\parallel}$ around $\omega_{\parallel}=-1$. The rest of the parameter values are: $\omega_{\perp}=3$, $\Omega_{\mathrm{HF}}=50$ and $\varphi=\pi/2$. The analytical results obtained using the averaging method [i.e.,  Eq.~(\ref{simazavmsrnonzero}) for all the values of $\omega_{\parallel}$] are plotted with dashed lines, and those obtained from the multiple scale method [i.e., Eq.~(\ref{simazavmsrnonzero}) for  $\omega_{\parallel}\neq -1$, and Eq.~(\ref{msexpvalsigzrzero}) for $\omega_{\parallel}=-1$] with solid lines.   We also show with symbols the results obtained numerically. In the upper panel, it is considered the case $\omega_{\parallel}=-1.1$. In this case, the value of $\omega_{\parallel}$ is not too close to the resonant value $\omega_{\parallel}=-1$, and both the averaging and the multiple scale methods reproduce properly the numerical results depicted by crosses. In the middle panel, it is considered the cases $\omega_{\parallel}=-1.005$ (pluses), $\omega_{\parallel}=-1.001$ (circles) and $\omega_{\parallel}=-1.0005$ (triangles). As  $\omega_{\parallel}$ gets closer to the resonant value $\omega_{\parallel}=-1$, the disagreement between the analytical results in Eq.~(\ref{simazavmsrnonzero}) and the numerical ones becomes more pronounced, and the numerical results converge gradually to the multiple scale solution in Eq.~(\ref{msexpvalsigzrzero}). Since the averaging and the multiple scale solutions coincide when $\omega_{\parallel}\neq -1$ [see Eq.~(\ref{simazavmsrnonzero})], the dashed lines are hidden by the solid lines in the upper and middle panels. Finally, in the lower panel it is shown the case $\omega_{\parallel}=-1$. In this case, only the multiple scale expression in Eq.~(\ref{msexpvalsigzrzero}) is able to reproduce the numerical results depicted by squares. To calculate the multiple scale frequency in Eq.~(\ref{msfrequencyrzero}), we have used the value $\gamma_{1,1}\approx-0.603984$, which has been obtained numerically from Eq.~(\ref{defgamma1j}) using the function \texttt{NIntegrate} of the program \textsc{Mathematica}.    

\begin{figure}[htb]
\vspace{0.5cm}
\includegraphics[scale=0.29]{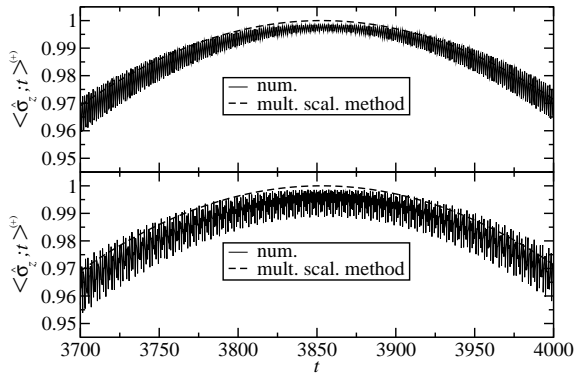}
\caption{Time dependence of $\langle \hat{\sigma}_z;t\rangle^{(+)}$ for $\omega_{\parallel}=-1$, $\omega_{\perp}=3$, $\Omega_{\mathrm{HF}}=50$, $r=r_1\approx 2.40483$ [first zero of the Bessel function $J_0(r)$], $\varphi=\pi/2$ (upper panel) and $\varphi=0$ (lower panel). Dashed line: analytical results obtained by the multiple scale method  [Eq.~(\ref{msexpvalsigzrzero})]. Solid lines: numerical results.} 
\label{timebehavior3}
\end{figure}

We finish this section with a brief discussion of the role played  by the phase shift $\varphi$ between  $\mathbf{B}_{\mathrm{HF}}(t)$ and $\mathbf{B}(t)$ in the system dynamics. Since the frequencies  $\Omega_{\mathrm{eff}}$, $\Omega_{\mathrm{ms}}$ and $\Omega_{\mathrm{ms},j}$ do not depend on $\varphi$,  the analytical results obtained up to now in this section are also independent of this parameter. As we shall see below, this is so due to the initial conditions  considered, namely $|\Psi,0\rangle=|\pm \rangle$. In the numerical results, the value of $\varphi$ affects only slightly the high-frequency and small-amplitude oscillations observed in $\langle \hat{\sigma}_z;t\rangle^{(\pm)}$ (see Fig.~\ref{timebehavior3}).

A more interesting situation arises when one considers an arbitrary initial condition of the form
\begin{equation}
\label{IC1}
|\Psi,0\rangle=c \,|+\rangle +e^{i \phi} \sqrt{1-c^2}\,|-\rangle,
\end{equation}
where $c\in [0,1]$ and $\phi\in[0,2 \pi)$. In this case, it is straightforward to see from Eqs.~(\ref{MSsolution}) and (\ref{expforexpvalsigz}) that, if we write explicitly  the dependence of the expected value of $\hat{\sigma}_z$ on the parameters $\varphi$ and $\phi$, it results that 
\begin{equation}
\label{phaseeffect}
\langle \hat{\sigma}_z;t,\{\varphi,\phi\}\rangle_{\mathrm{ms}}=\langle \hat{\sigma}_z;t,\{0,\phi^{\prime}\}\rangle_{\mathrm{ms}}\,,
\end{equation} 
where $\phi^{\prime}=\phi+r\sin\varphi$. That is, in order to calculate the expected value of $\hat{\sigma}_z$ by the multiple scale method, a high-frequency magnetic field with a phase shift $\varphi$ can be replaced by another one with a phase shift equal to $0$, if the phase $\phi$ of initial state  is also replaced by another value $\phi^{\prime}=\phi+r\sin\varphi$. From Eqs.~(\ref{AMsolution2}) and (\ref{expforexpvalsigz}), it is easy to see that this result also holds if one uses the averaging method.

As an example of this property, let us consider the case in which $r=r_j$ and $\omega_{\parallel}=-1$. For these parameter values, the effective Hamiltonian in Eq.~(\ref{effHamiltonian}) vanishes and, consequently, from Eqs.~(\ref{AMsolution2}), (\ref{expforexpvalsigz}) and (\ref{IC1}), the averaging method yields 
\begin{equation}
\label{avarbitraryIS}
\langle \hat{\sigma}_z;t\rangle_{\mathrm{av}}=2 c^2 -1,
\end{equation}
which is independent of $\varphi$ and $\phi$. By contrast, if one uses the multiple scale method, from Eqs.~(\ref{MSHamiltonian0}), (\ref{MSsolution}), (\ref{expforexpvalsigz}) and (\ref{IC1}), one obtains
\begin{eqnarray}
\label{msarbitraryIS}
\langle \hat{\sigma}_z;t\rangle_{\mathrm{ms}}\!\!\!\!&=&\!\!\!\!(2 c^2-1)\cos\left(\Omega_{\mathrm{ms},j} t\right)\nonumber\\
\!\!\!\!&&\!\!\!\!-\,2c\sqrt{1-c^2}\sin\left(\phi^{\prime}\right)\sin\left(\Omega_{\mathrm{ms},j} t\right),\nonumber\\
\!\!\!\!&&\!\!\!\!
\end{eqnarray}
which depends on $\varphi$ and $\phi$ through $\phi^{\prime}$.

\begin{figure}[htb]
\includegraphics[scale=0.29]{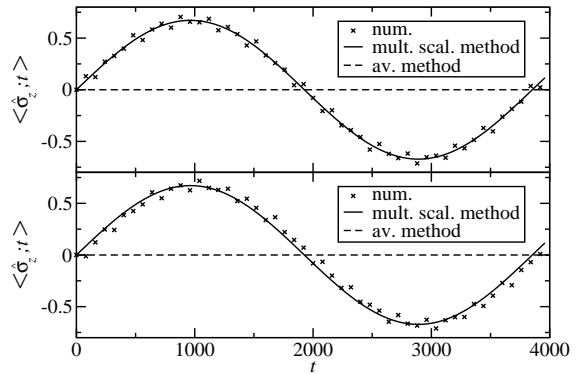}
\caption{Time dependence of $\langle \hat{\sigma}_z;t\rangle$ for $\omega_{\parallel}=-1$, $\omega_{\perp}=3$, $\Omega_{\mathrm{HF}}=50$, $r=r_1\approx 2.40483$ [first zero of the Bessel function $J_0(r)$] and an initial condition of the form (\ref{IC1}) with $c=2^{-1/2}$. Upper panel: $\varphi=\pi/2$ and $\phi=0$. Lower panel: $\varphi=0$ and $\phi=r_1$. Dashed lines: analytical results obtained from Eq.~(\ref{avarbitraryIS}) (averaging method). Solid lines: analytical results obtained from Eq.~(\ref{msarbitraryIS}) (multiple scale method). Crosses: numerical results.}
\label{timebehavior4}
\end{figure}

In order to compare these analytical expressions with the numerical results, in Fig.~\ref{timebehavior4} it is shown the time dependence of $\langle \hat{\sigma}_z;t\rangle$ for the following parameter values: $\omega_{\parallel}=-1$, $\omega_{\perp}=3$, $\Omega_{\mathrm{HF}}=50$ and $r=r_1\approx 2.40483$. In the upper panel, it is considered a phase shift $\varphi=\pi/2$ and an initial condition of the form (\ref{IC1}) with $c=2^{-1/2}$ and $\phi=0$.  In the lower panel, it is assumed a phase shift $\varphi=0$ and the same initial condition as before but with $\phi=r_1$. According to Eq.~(\ref{msarbitraryIS}), the  multiple scale method leads to the same conclusion in both panels (solid lines). The analytical results obtained using the averaging method [Eq.~(\ref{avarbitraryIS})] are plotted with dashed lines. We also show with crosses the results obtained numerically. Once again, only the multiple scale method is able to  reproduce the numerical results, with exception of the high-frequency and small-amplitude oscillations.

\section{Conclusions}
\label{Conclusions}
We have studied how the resonant behavior displayed by a spin-$1/2$ particle under the influence of a rotating magnetic field is affected by the application of an additional high-frequency magnetic field along the rotation axis. Analytical expressions for the time evolution operator and the expected value of $\hat{\sigma}_z$ have been obtained by using two alternative methods, namely, the averaging method and the multiple scale method. 

When the ratio of the strength of the high-frequency magnetic field to its frequency (in dimensionless units) is not a zero of the Bessel function $J_0(r)$, both methods lead to the same conclusion about the effect of the high-frequency magnetic field on the resonant behavior. Specifically, its presence only gives rise to a narrowing of the resonant peak. By contrast, when this ratio coincides with a zero of $J_0(r)$, both methods lead to different predictions. More precisely, the averaging method predicts the complete disappearance of the resonant behavior, whereas according to the multiple scale method the resonant peak persists but its width becomes zero. This discrepancy in the conclusions of both methods is similar to the one appearing in the study of the phenomenon of coherent destruction of tunneling. There, the averaging method predicts the complete destruction of tunneling, whereas other more accurate methods, among them the multiple scale method, conclude that this destruction is only temporary. In our case, the numerical results show up the existence of an extremely narrow resonant peak and, consequently, confirm the prediction made by the multiple scale method. An open problem which arises in this context is how to extend the multiple scale method proposed in this paper in order to explain the finite width of the resonant peak observed in the numerical solution. 

Comparison of the analytical results for the expected value of $\hat{\sigma}_z$ with the numerical ones shows that the multiple scale method also allows a better description of the time evolution than the averaging method, above all in the long-time limit.

Finally, we have studied the role played by the phase shift between the high-frequency and the rotating magnetic fields in the system dynamics. From this study it follows that a modification of this phase shift is basically equivalent to a change in the phase of the initial state.   

\section{Acknowledgements}

This paper is dedicated to Professor Peter H\"anggi on the occasion of his 60th birthday. The author acknowledges the support of the Ministerio de Ciencia e Innovaci\'on of Spain (FIS2008-02873) and the Junta de Andaluc\'{\i}a.






\begin{thebibliography}{00}
\bibitem{nota1} We assume that $\omega_{\perp}\neq 0$, because otherwise the solution of Eq.~(\ref{evolutionoperator1}) is trivial.
\end{thebibliography}



\end{document}